\documentclass[11pt]{article}

\usepackage[margin=1in,footskip=0.25in]{geometry}

\usepackage{graphicx}
\usepackage{amsmath,amssymb, color}
\usepackage{amsfonts}
\usepackage{subfigure}
\usepackage{longtable}
\usepackage{multirow}
\usepackage{verbatim}
\usepackage{setspace}
\usepackage{tabularx, booktabs, threeparttable}
\usepackage{pdflscape}

\usepackage{tikz}
\usetikzlibrary{shapes}

\usepackage{natbib}

\usepackage{amsmath}

\usepackage{times}

\usepackage{bm}

\usepackage[plain,noend]{algorithm2e}

\makeatletter
\renewcommand{\algocf@captiontext}[2]{#1\algocf@typo. \AlCapFnt{}#2} 
\def\@algocf@capt@plain{top}
\renewcommand{\algocf@makecaption}[2]{%
  \addtolength{\hsize}{\algomargin}%
  \sbox\@tempboxa{\algocf@captiontext{#1}{#2}}%
  \ifdim\wd\@tempboxa >\hsize
    \hskip .5\algomargin%
    \parbox[t]{\hsize}{\algocf@captiontext{#1}{#2}}
  \else%
    \global\@minipagefalse%
    \hbox to\hsize{\box\@tempboxa}
  \fi%
  \addtolength{\hsize}{-\algomargin}%
}
\makeatother

\newcommand{\vb}{{b}}

\newcommand{\vk}{{k}}

\newcommand{\vr}{{r}}

\newcommand{\vx}{{x}}

\newcommand{\vz}{{z}}

\newcommand{\vX}{{X}}

\newcommand{\veins}{{1}}
\newcommand{\vnull}{{0}}

\newcommand{\vdelta}{\mbox{$\delta$}}

\newcommand{\vbeta}{\mbox{$\beta$}}

\newcommand{\bay}{\begin{array}}
\newcommand{\eay}{\end{array}}

\newcommand{\ra}{\rightarrow}

\newcommand{\bqa}{\begin{eqnarray*}}
\newcommand{\eqa}{\end{eqnarray*}}
\newcommand{\bqan}{\begin{eqnarray}}
\newcommand{\eqan}{\end{eqnarray}}
\newcommand{\bqt}{\begin{quote}}
\newcommand{\eqt}{\end{quote}}
\newcommand{\bt}{\begin{tabbing}}
\newcommand{\et}{\end{tabbing}}
\newcommand{\bit}{\begin{itemize}}
\newcommand{\eit}{\end{itemize}}
\newcommand{\ben}{\begin{enumerate}}
\newcommand{\een}{\end{enumerate}}
\newcommand{\beq}{\begin{equation}}
\newcommand{\eeq}{\end{equation}}
\newcommand{\bdefi}{\begin{definition}}
\newcommand{\edefi}{\end{definition}}
\newcommand{\bpro}{\begin{proposition}}
\newcommand{\epro}{\end{proposition}}
\newcommand{\blem}{\begin{lemma}}
\newcommand{\elem}{\end{lemma}}

\newcommand{\bco}{\begin{corollary}}
\newcommand{\eco}{\end{corollary}}
\newcommand{\bdes}{\begin{description}}
\newcommand{\edes}{\end{description}}

\newtheorem{definition}{Definition}[section]
\newtheorem{proposition}[definition]{Proposition}
\newtheorem{lemma}[definition]{Lemma}
\newtheorem{theorem}[definition]{Theorem}
\newtheorem{corollary}[definition]{Corollary}
\newtheorem{remark}{Remark}

\DeclareFontFamily{U}{mathx}{\hyphenchar\font45}
\DeclareFontShape{U}{mathx}{m}{n}{
      <5> <6> <7> <8> <9> <10>
      <10.95> <12> <14.4> <17.28> <20.74> <24.88>
      mathx10
      }{}
\DeclareSymbolFont{mathx}{U}{mathx}{m}{n}
\DeclareFontSubstitution{U}{mathx}{m}{n}
\DeclareMathAccent{\widecheck}{0}{mathx}{"71}

\def\boxit#1{\vbox{\hrule\hbox{\vrule\kern6pt
          \vbox{\kern6pt#1\kern6pt}\kern6pt\vrule}\hrule}}

\def\bfblue#1{{\color{blue}\bf#1}}

\DeclareMathOperator*{\argmin}{arg\,min}

\newcommand{\symbolALg}{\raisebox{0.5pt}{\tikz{\node[draw,scale=0.4,circle,fill=none](){};}}}
\newcommand{\symbolALgg}{+}
\newcommand{\symbolLa}{\raisebox{0.5pt}{\tikz{\node[draw,scale=0.3,regular polygon, regular polygon sides=3,fill=none,rotate=0](){};}}}
\newcommand{\symbolLn}{\(\times\)}

 \author{ Lan Wang \thanks{Research supported by the U.S.~National Science Foundation.} 
 \\ School of Statistics \\ University of Minnesota \\ \texttt{wangx346@umn.edu} \\[20pt]
  \and Ingrid Van Keilegom
  \thanks{Research supported by the European Research Council and an Interuniversity Attraction Pole research network of the Belgian government.}\\ Research Centre for Operations Research and Business Statistics\\  KU Leuven\\ \texttt{ingrid.vankeilegom@kuleuven.be} \\[20pt]
     \and Adam Maidman \footnotemark[1]
  \\ School of
  Statistics\\ University of Minnesota\\ \texttt{maidm004@umn.edu} }
\title{Wild Residual Bootstrap Inference for Penalized Quantile Regression with Heteroscedastic Errors} \date{\today}

\allowdisplaybreaks

\begin{document}

\maketitle

\begin{abstract}
We consider a heteroscedastic regression model in which some of the regression coefficients 
are zero but it is not known which ones. 
Penalized quantile regression is a useful approach for analyzing such data.
By allowing different covariates to be relevant for modeling conditional quantile functions at different quantile levels,
it provides a more complete picture of the
conditional distribution of a response variable than mean regression. 
Existing work on penalized quantile regression has been mostly focused on point estimation. 
Although bootstrap procedures have recently been shown to be effective for inference
for penalized mean regression,
they are not directly applicable to penalized quantile regression with heteroscedastic errors. 
We prove that a wild residual bootstrap procedure 
for unpenalized quantile regression  is asymptotically valid for approximating the distribution of a
penalized quantile regression estimator with an adaptive $L_1$ penalty and that
a modified version can be used to approximate the distribution of $L_1$-penalized
quantile regression estimator. The new methods do not need to 
estimate the unknown error density function. We establish consistency,
demonstrate finite sample performance, and  illustrate the applications on a real data example.
\end{abstract}


\section{Introduction}

We consider the quantile regression model 
$Y_i=\vx_i^T\vbeta_0+\epsilon_i$ ($i=1, \ldots, n$),
where $\vx_i=(x_{i0},x_{i1},\ldots,x_{ip})^T$ with $x_{i0}=1$ is the $i$th nonstochastic design point in $\mathcal{R}^p$, 
and $\epsilon_i$ is a random error with probability density $f_i$ and the $\tau$th quantile equal to zero.
The unknown regression coefficient $\vbeta_0=(\beta_{00},\beta_{01},\ldots,\beta_{0p})^T$ may depend on $\tau$, but we omit such dependence
in notation for simplicity.
Quantile regression was proposed by \cite{K4} and has become a popular alternative
to least squares regression.  
Conditional quantiles are of interest in a variety of applications, such as the 
conditional median of medical expenditure or a low conditional quantile of birth weight.
Comparing such quantiles for a range of $\tau$ values enables researchers to obtain a more complete picture of
the conditional distribution than mean regression and is particularly useful for analyzing heterogeneous data. 
See \cite{K2} and \cite{K3}.

We suppose that some of the covariates 
are irrelevant for modeling the $\tau$th conditional quantile but we have no prior information 
on which.
In such a setting, penalized quantile regression has been proven to avoid over-fitting
by shrinking the estimated coefficients of irrelevant covariates
toward zero. Here, we focus on the asymptotic regime where the number of predictors $p$ is fixed
while the sample size $n$ goes to infinity.
Asymptotic theory for penalized quantile regression in this setup was recently studied by
\cite{Z4} for independent and identically distributed random errors,
and \cite{W2}, who established the 
asymptotic distribution of penalized quantile regression estimator for 
the adaptive $L_1$ penalty \citep{Z2}
and considered an extension to the general heteroscedastic error setting.
However, these works have not considered estimation of the standard error of the estimated penalized quantile regression 
coefficients. The asymptotic distribution of 
$L_1$-penalized quantile regression has 
a positive probability mass at zero
for the component for which the true regression parameter has a zero value.
Inference based directly on asymptotic theory is not convenient.
On the other hand,
the adaptively $L_1$-penalized quantile regression estimator enjoys the oracle property under regularity conditions:
the zero coefficients are estimated as exactly zero with probability approaching unity and
the nonzero coefficients have the asymptotic normal distribution we would obtain if we knew
in advance which coefficients are zero. However, convergence to the oracle distribution is often slow and results
in inaccurate confidence intervals \citep{C3}.  

In practice, a two-step procedure is commonly used to construct confidence intervals. First,
penalized quantile regression is applied to select variables. Then the model is refitted with
selected variables only to construct confidence intervals. 
Such a procedure does not account for uncertainties involved in variable selection and generally
tends to produce wider confidence intervals, as demonstrated in our simulation study.

These challenges motivate us to develop a wild residual bootstrap-based inference 
approach for penalized quantile regression 
with $L_1$ or adaptive $L_1$ penalty. 
Our work is mostly related to 
\cite{C1,C2,C3}
 and \cite{Ca}
on bootstrapping penalized estimators in the least squares regression
setting.
An alternative perturbation method for inference on regularized regression estimates was studied in \citet{Minnier2011}.
\cite{C1}
proved that standard bootstrap
is inconsistent for estimating the distribution of the $L_1$ penalized least squares estimator
when one or more of the components of  the 
regression parameter vector are zero; the failure of the naive paired bootstrap was proved in \cite{Ca}.
Modified residual and paired bootstraps were proposed in \cite{C2}
and \cite{Ca}, respectively.
\cite{C3} demonstrated that although the adaptively penalized
least squares estimator enjoys the oracle property, inference based directly on the oracle
distribution is often inaccurate and more accurate inference can be obtained via a residual bootstrap. 
However, these bootstrap methods do not directly apply to the quantile regression setting due to
the nonsmoothness of the quantile loss function and the heteroscedastic error distribution.
We prove that a wild residual bootstrap procedure proposed by \cite{F1}
for unpenalized quantile regression  is asymptotically valid for approximating the distribution of
the quantile regression estimator with adaptive $L_1$ penalty. Furthermore,
a modified version of this wild residual bootstrap procedure 
can be used to approximate the distribution of $L_1$ penalized
quantile regression. Our derivation of the bootstrap consistency theory for 
penalized quantile regression uses techniques substantially different from that of \cite{F1}.

\section{Inference for adaptive $L_1$-penalized quantile regression}
\subsection{Quantile regression with adaptive $L_1$ penalty}

The unpenalized quantile regression estimator for $\vbeta_0$ is
$\overline{\vbeta}=(\overline{\beta}_0, \ldots, \overline{\beta}_p)^T$,
where
\bqan \label{oqreg}
\overline{\vbeta}=\argmin_{\vbeta}\sum_{i=1}^n\rho_{\tau}(Y_i-\vx_i^T\vbeta)
\eqan
and
$\rho_{\tau}(u) = u\left\{\tau - I(u<0)\right\}$ is the quantile
loss function. Under general regularity conditions,
$\overline{\beta}$ is asymptotically normal. The asymptotic covariance matrix of $\overline{\beta}$ 
depends on the unknown
conditional density function of $\epsilon_i$ \citep{K2}.

Often not all covariates collected are relevant for modeling the
$\tau$th conditional quantile, that is, some of the components of $\vbeta_0$ are zero. 
Let $A=\{1\leq j\leq p: \beta_{0j}\neq 0\}$ be the index set of the nonzero coefficients.
Let $|A|=q$ be the cardinality of the set $A$. 
Without loss of
generality, we assume that the last $p-q$ components of $\vbeta_0$ are
zero; that is, we can write $\vbeta_0=(\vbeta_{01}^T,\vnull^T_{p-q})^T$, where $\vnull_{p-q}$
denotes a $(p-q)$- dimensional vector of zeros, and $A=\{1, \ldots, q\}$.
Let $\vX=(\vx_1,\ldots,\vx_n)^T$ be the $n\times (p+1)$ matrix of
covariates, where $\vx_1^T,\ldots,\vx_n^T$ are the rows of $\vX$. We also write
$X=(\veins, \vX_1,\ldots,\vX_{p})$, where $\veins,\vX_1,\ldots,\vX_p$
are the columns of $\vX$ and $\veins$ represents an $n$-vector of ones.
Define $\vX_{A}$ to be the submatrix of $\vX$ that consists of its first $q+1$
columns; and define $\vX_{A^c}$ to be the submatrix of $\vX$ that
consists of its last $p-q$ columns.  Similarly, let $\vx_{iA}$ be the subvector that contains 
the first $q+1$ entries of $\vx_{i}$.

The quantile regression estimator with the adaptive $L_1$ penalty performs simultaneous estimation and variable 
selection by minimizing a penalized quantile loss function, i.e.,
\bqan \label{pqreg}
\widetilde{\vbeta}=\argmin_{\vbeta}
\Big\{\sum_{i=1}^n\rho_{\tau}(Y_i-\vx_i^T\vbeta)+\lambda_n\sum_{j=1}^{p}w_j|\beta_j|\Big\},
\eqan 
where $\lambda_n>0$ is a tuning parameter, and
$w_j=|\overline{\beta}_j|^{-\gamma}$ are the adaptive weights 
($\gamma>0$).
Write $\widetilde{\vbeta}=(\widetilde{\beta}_0, \ldots, \widetilde{\beta}_p)^T$
and $\widetilde{A}=\{1 \le j \le p: \widetilde{\beta}_j\neq 0\}$.
Let $\widetilde{\vbeta}_1$ be the subvector that contains the first $(q+1)$ elements of $\widetilde{\vbeta}$.
Let $D_0=\lim_{n\ra \infty}n^{-1}\sum_{i=1}^n\vx_{iA}\vx_{iA}^T$
and $D_1=\lim_{n\ra \infty}n^{-1}\sum_{i=1}^nf_i(0)\vx_{iA}\vx_{iA}^T$, where $f_i(0)$
is the  
density function of $\epsilon_i$ 
evaluated at zero.
The following properties of $\widetilde{\vbeta}$ were established in \cite{W2}. 

\blem\label{liona}
Assume Condition 2 of Section 2.2 is satisfied. 
If  $n^{-1/2}\lambda_n\ra 0$ and $n^{(\gamma-1)/2}\lambda_n\ra \infty$, then
the adaptive $L_1$-penalized quantile regression estimator $\widetilde{\vbeta}$ enjoys the oracle property. That is,\\
(i) $\mbox{pr}(\widetilde{A}=A)\ra 1$ as $n\ra \infty$;\\
(ii) $n^{1/2}(\widetilde{\vbeta}_{1}-\vbeta_{01})\ra N\{\vnull_{q+1}, \tau(1-\tau)D_1^{-1}D_0D_1^{-1}\}$
in distribution as $n\ra \infty$.
\elem
The result in Lemma 1 is referred to as the oracle property: with probability approaching one
the zero coefficients of $\vbeta_0$ are identified as zero and the nonzero coefficients are identified as nonzero;
and we can estimate the nonzero subvector of $\vbeta_0$ as efficiently as if we know the true model in advance. 
The proof of Lemma \ref{liona} is given in the Supplementary Material.

\subsection{A wild residual bootstrap procedure and its consistency}
\label{Section22}

We use a wild residual bootstrap procedure to approximate the asymptotic distribution of  $\widetilde{\vbeta}$. 
Our procedure is 
motivated by the work of \cite{F1} for unpenalized quantile regression.
To obtain the wild bootstrap sample, we follow the steps below.
\begin{enumerate}
\item We first calculate the residuals from the adaptively penalized quantile regression:
$\hat{\epsilon}_i=Y_i-\vx_i^T\widetilde{\vbeta}$ ($i=1, \ldots, n$) and
obtain $\widetilde{\vbeta}$ by (\ref{pqreg}).
\item  Let $\epsilon_i^*=r_i|\hat{\epsilon}_i|$, where $r_i$ ($i=1,\ldots, n$) are generated as a random sample
from a distribution with a cumulative distribution function $G$ satisfying Conditions 3-5 below.
\item
We generate the bootstrap sample as
$Y_i^*=\vx_i^T\widetilde{\vbeta}+\epsilon_i^*$ ($i=1,\ldots,n$).
\end{enumerate}

Using the bootstrap sample, we recalculate the adaptively penalized quantile regression estimator
as
\bqan \label{btpqreg}
\widetilde{\vbeta}^*=\argmin_{\vbeta}\Big\{\sum_{i=1}^n\rho_{\tau}(Y_i^*-\vx_i^T\vbeta)
+\lambda_n\sum_{j=1}^{p}w_j^*|\beta_j|\Big\},
\eqan
where $w_j^*=|\overline{\beta}_j^*|^{-\gamma}$,
$\overline{\vbeta}^*=(\overline{\beta}^*_0,  \ldots, \overline{\beta}^*_p)^T$
is the ordinary quantile regression estimator recomputed on the bootstrap sample.
For $j=1, \ldots, p$ and $0<\alpha<1$, let $d_j^{*(\alpha/2)}$ and $d_j^{*(1-\alpha/2)}$ be the $(\alpha/2)$-th and 
$(1-\alpha/2)$-th quantiles of the
bootstrap distribution of   $n^{1/2}(\widetilde{\vbeta}_j^*-\widetilde{\vbeta}_j)$, respectively. 
We can estimate $d_j^{*(\alpha/2)}$ and $d_j^{*(1-\alpha/2)}$ from a large number of bootstrap samples.
An asymptotic $100(1-\alpha)\%$
bootstrap confidence interval for $\beta_{0j}$,   $j=1,\ldots, p$, is given by
$
\big[\widetilde{\vbeta}_j-n^{-1/2}d_j^{*(1-\alpha/2)}, \widetilde{\vbeta}_j-n^{-1/2}d_j^{*(\alpha/2)}\big].
$
As in \cite{F1}, we work under the following technical conditions:\\[.3cm]
Condition 1.   The true value $\vbeta_0$ is an interior point of a compact set in
$R^p$.
The density of $\epsilon_i$,
denoted by $f_i(\cdot)$, is Lipschitz continuous and is bounded away from 0 and
$\infty$ in a neighborhood around 0 for all $i$;\\
Condition 2. 
$\lim_{n\ra \infty}n^{-1}\sum_{i=1}^n\vx_{i}\vx_{i}^T\ra B_0$
and $\lim_{n\ra \infty}n^{-1}\sum_{i=1}^nf_i(0)\vx_{i}\vx_{i}^T\ra B_1$ for some positive definite matrices $B_0$
and $B_1$.  Furthermore,
$\sum_{i=1}^n||\vx_i||^3=O(n)$ and $\max_{1 \le i \leq n}||\vx_i||=O(n^{1/4})$, where $||\cdot||$ is the Euclidean norm;\\
Condition 3.   for some strictly positive constants $c_1$ and $c_2$, 
$\sup\{r\in \mathbb{G}: r\leq 0\}=-c_1$ and 
$\inf\{r\in \mathbb{G}: r\geq 0\}=c_2$, where $\mathbb{G}$ is the support of the weight distribution $G$;\\
Condition 4.  the weight distribution $G$ satisfies $\int_{0}^{+\infty}r^{-1}dG(r)=-\int_{-\infty}^0r^{-1}dG(r)=1/2$
and $E_{G}(|r|)<\infty$ , where the expectation is taken under $G$;\\
Condition 5.  the $\tau$th quantile of the distribution $G$ is zero.\\

Theorem \ref{Balasso} shows that the conditional distribution of 
$n^{1/2}(\widetilde{\vbeta}^*-\widetilde{\vbeta})$
provides an asymptotically valid approximation of that of $n^{1/2}(\widetilde{\vbeta}-\vbeta)$. 
Let $\widetilde{A}^*=\{j=1,\ldots, p: \widetilde{\beta}_j^*\neq 0\}$, and let $\widetilde{\vbeta}_1^*$ be the subvector that contains the first $q+1$ elements of $\widetilde{\vbeta}^*$.
Let $\vr=\{r_1, \ldots, r_n\}$ be the random bootstrap weights and $\vz=\{\vz_1, \ldots, \vz_n\}$ be the random sample.
By the wild bootstrap mechanism, the distribution of $\vr$ is independent of that of $\vz$.
Let $\mbox{pr}_{\vz}$ denote the probability under the joint distribution of $\vz$,
and let $\mbox{pr}_{\vr\mid \vz}$ denote the probability of $\vr$
conditional on $\vz$. 

\begin{theorem}\label{Balasso} If Conditions 1--5 and the assumptions of Lemma \ref{liona} are satisfied, 
then $pr_{\vr\mid \vz}(\widetilde{A}^*=A)=1+o_{ \textnormal{pr}_{\vz}}(1)$. Furthermore,
\bqa
\sup_{t}\big| \textnormal{pr}_{\vr\mid \vz}\{n^{1/2}(\widetilde{\vbeta}_1^*-\widetilde{\vbeta}_{1})\leq t\}-
 \textnormal{pr}_{\vz}\{n^{1/2}(\widetilde{\vbeta}_{1}-\vbeta_{01})\leq t\}\big|=o_{ \textnormal{pr}_{\vz}}(1).
\eqa
\end{theorem}

\begin{remark} Conditions 1 and 2 are slightly weaker than the corresponding conditions in \cite{F1}.
Under  Condition 5, conditional on the data, $\epsilon_i^*$ has the $\tau$th quantile equal
to zero.  Conditions 3 and 4 ensure that the asymptotic distribution of the bootstrap estimator, conditional
on the data, matches the unconditional asymptotic distribution of the original adaptively penalized quantile regression estimator,
which depends on the unknown error density function.
 A simple weight distribution that satisfies Conditions 3--5 is the two-point distribution with probabilities
$1-\tau$ and $\tau$ at $r=2(1-\tau)$ and $-2\tau$, respectively. Another example 
given in  \cite{F1} is the distribution
which for $1/8<\tau<7/8$,
$
g(r)=G'(r)=-rI(-2\tau-1/4\leq r \leq -2\tau+1/4)+rI\{2(1-\tau)-1/4\leq r \leq 2(1-\tau)+1/4\}.
$
We propose several other distributions that satisfy these conditions in the Supplementary Material.
\end{remark}

\begin{remark} By definition $n^{1/2}(\widetilde{\vbeta}^*-\widetilde{\vbeta})$ minimizes $Q_n^*(\vdelta)$,
where
$
Q_n^*(\vdelta)=\sum_{i=1}^n\big\{\rho_{\tau}(\epsilon_i^*-n^{-1/2}\vx_i^T\vdelta)-\rho_{\tau}(\epsilon_i^*)\big\}
+\lambda_n\sum_{j=1}^pw_j^*\big(\widetilde{\beta}_j+n^{-1/2}\delta_j|-|\widetilde{\beta}_j|\big),
$
The crux of the proof of Theorem \ref{Balasso} is to show that conditional on the data,
\bqa
Q_n^*(\vdelta)\ra Q^*(\vdelta)=
\begin{cases}
-\vdelta^TH+\vdelta^TB_1\vdelta/2, &\text{ $\delta_j=0$ for $j>q$},\\
+\infty, &\text{otherwise},
\end{cases}
\eqa
in probability, where  $H\sim N\{\vnull, \tau(1-\tau)B_0\}$. Then the results follow from epi-convergence theory, 
see the unpublished technical reports of Geyer ({\it On the asymptotics of
convex stochastic optimization}, technical report, 1996) and Knight ({\it Epi-convergence in distribution and stochastic
equi-semicontinuity}, technical report, 1999).
\end{remark}

\begin{remark}
As pointed out by a referee, \citet{Leeb2008} and \citet{Potscher2009} revealed that
the distribution of adaptive lasso and other shrinkage-type estimators cannot be estimated 
uniformly in a shrinking neighborhood of the underlying parameter values. 
In the setting we consider, the number of covariates is fixed. We assume the smallest nonzero signal
is  not diminishing to zero when the sample size increases. Furthermore,
as in \cite{C2}, we do not claim the bootstrap based estimator of the distribution of  adaptive lasso to be uniformly
consistent over any diminishing neighborhood of underlying parameter values. See also Remark 3 of  \cite{C2}.
\end{remark}

\begin{remark}
For the adaptive lasso, the coverage probability of the confidence interval approaches 
unity, because the wild residual bootstrap distribution approximates the 
adaptive lasso estimator distribution, which identifies zero coefficients as exactly
zero with probability approaching unity. 
\end{remark}

\section{Modified wild residual bootstrap for $L_1$ penalized quantile regression}
We also consider the $L_1$ or lasso penalized quantile regression estimator
\bqan \label{qlasso}
\widecheck{\vbeta}=\argmin_{\vbeta}\Big\{\sum_{i=1}^n\rho_{\tau}(Y_i-\vx_i^T\vbeta)
+\lambda_n\sum_{j=1}^{p}|\beta_j|\Big\},
\eqan 
where $\lambda_n>0$ is a tuning parameter.
The asymptotic distribution of $\widecheck{\vbeta}$ follows that of the minimizer of 
a random process, which is specified in the following lemma.
\blem\label{bird}
Under Condition 2 and
if  $n^{-1/2}\lambda_n\ra \lambda_0 \ge 0$, 
\bqa
n^{1/2}(\widecheck{\vbeta}-\vbeta_0)&\ra &\argmin_{\vdelta}\Big[-\vdelta^TH+\vdelta^TB_1\vdelta/2
+\lambda_0\sum_{j=1}^p\big\{|\delta_j|I(\beta_{0j}=0)\\
&&\hspace{2cm}+\delta_j\mathrm{sign}(\beta_{0j})I(\beta_{0j}\neq 0)\big\}\Big],
\eqa
in distribution as $n\ra \infty$, where $H$ is defined in Remark 2.
\elem
The proof is given in the Supplementary Material.
For $L_1$-penalized mean regression, \cite{C1} proved that the asymptotic distribution
of the naive residual bootstrapped lasso estimator is a random measure on $R^p$ and that the bootstrap is inconsistent
whenever the regression parameter vector contains one or more zeros. 
An explanation of this phenomenon is that the lasso estimates the sign of nonzero coefficients correctly with high
probability, but estimates the zero coefficients to be positive or negative with positive probabilities. The naive residual
bootstrap fails to reproduce the sign of zero coefficients with high probability. To remedy this,  
\cite{C1} proposed a thresholding procedure, which we adapt.

Our procedure proceeds as follows.
Let $\{a_n\}$ be a sequence of numbers such that $a_n+(n^{-1/2}\log n)a_n^{-1}\ra 0$
as $n\ra \infty$. For example, $a_n=cn^{-\delta}$, for some $c>0$, $0<\delta<1/2$.
For $\overline{\vbeta}$ defined in (\ref{oqreg}),
we consider the thresholded estimator $\widecheck{\vbeta}^*=(\widecheck{\beta}^*_0, \ldots, \widecheck{\beta}^*_p)^T$,
where $\widecheck{\beta}^*_0=\overline{\beta}_0$ and $\widecheck{\beta}^*_j=\widecheck{\beta}_jI(|\widecheck{\beta}_j|>a_n)$ for $j=1\ldots,p$.
Let $\widecheck{\epsilon}_i=Y_i-\vx_i^T\widecheck{\vbeta}^*$ ($i=1, \ldots, n$).
Let $\epsilon_i^{**}=r_i|\widecheck{\epsilon}_i|$ ($i=1, \ldots, n$), where 
the bootstrap weights $r_i$ satisfy Conditions 3--5. 
We choose to threshold the ordinary quantile regression estimator directly.
Alternatively, we may threshold the lasso estimator $\widecheck{\vbeta}$, which will yield the same asymptotic results
for the bootstrapped estimator but requires an additional tuning parameter for the lasso.

The bootstrap sample is generated by
$Y_i^{**}=\vx_i^T\widecheck{\vbeta}^*+\epsilon_i^{**}$ ($i=1,\ldots,n$).
We then recalculate the $L_1$ penalized quantile regression estimator using the bootstrap sample:
\bqan \label{btqlasso}
\widecheck{\vbeta}^{**}=\argmin_{\vbeta}\Big\{\sum_{i=1}^n\rho_{\tau}(Y_i^{**}-\vx_i^T\vbeta)
+\lambda_n\sum_{j=1}^{p}|\beta_j|\Big\}.
\eqan
Theorem \ref{Bqlasso} below shows that the conditional distribution of 
$n^{1/2}(\widecheck{\vbeta}^{**}-\widecheck{\vbeta}^*)$
provides an asymptotically valid approximation of that of $n^{1/2}(\widecheck{\vbeta}-\vbeta_0)$.

\begin{theorem}\label{Bqlasso} 
If Conditions 1--5 and the assumptions of Lemma \ref{bird} are satisfied, then
\bqa
\sup_{t}\big|\textnormal{pr}_{\vr\mid \vz}\{n^{1/2}(\widecheck{\vbeta}^{**}-\widecheck{\vbeta}^*)\leq t\}-
\textnormal{pr}_{\vz}\{n^{1/2}(\widecheck{\vbeta}-\vbeta_0)\leq t\}\big|=o_{\textnormal{pr}_{\vz}}(1).
\eqa
\end{theorem}

\section{Numerical results}

\subsection{Monte Carlo studies}

We study the accuracy of 95\% confidence intervals constructed by 
our bootstrap procedures.
For the adaptive $L_1$ penalty, we select the tuning parameter $\lambda_n$ by minimizing a Bayesian information criterion   
\citep{Lee2014} and consider $\gamma=1,2$.
For the $L_1$ penalty, we select $\lambda_n$ by cross-validation and consider two choices
of $a_n$. One choice adopts a data-driven approach
that minimizes the estimated mean
squared error $E^*(||\widecheck{\vbeta}^{**}-\widecheck{\vbeta}^*||^2)$,
where $E^*$ is the average over bootstrap samples;
see Section 5.2 of \cite{C2} and Remark 2 of \cite{Ca}. The other choice is 
the empirical choice $a_n=n^{-1/3}$,
which is motivated by the rate required by the asymptotic theory.
The bootstrap random weights $r_i$ are generated from 
the two-point distribution described in \citet{F1}; see Remark 1.
We also tried alternative weight distributions and found the results similar. 

We compare the new methods with the confidence intervals from the oracle model, 
from the full model, and
from the two-step procedure described in Section 1 with adaptive lasso or lasso applied in the first step. 
The oracle procedure is not implementable in real data analysis. 
For these competing methods, 
we consider confidence intervals obtained by the rank score method 
and by the wild bootstrap method
in the R package quantreg \citep{quantreg}.

\begin{table}[ht]
\caption{Empirical coverage probabilities ($\times100$) and average interval lengths (in parentheses)
for nominal 95\% confidence intervals}
{\scalebox{0.9}{
\centering
\begin{tabular}{lcccccccc}
   \\
    & $\beta_1 = \Phi^{-1}(\tau)$ & $\beta_3 = $0$\cdot$25 & $\beta_5 = $0$\cdot$5 & $\beta_7 = 1$ & $\beta_9 = 2$ & Zeros & TP & FP \\[5pt]
  & & &   $ \tau=$0$\cdot$5 &  $n=100$ &  & & & \\[5pt]
   New AL1 & 92$\cdot$0 (0$\cdot$33) & 94$\cdot$6 (0$\cdot$15) & 93$\cdot$2 (0$\cdot$17) & 95$\cdot$3 (0$\cdot$13) & 92$\cdot$7 (0$\cdot$14) & 97$\cdot$4 (0$\cdot$06) & 4 & 0$\cdot$3 \\ 
  New AL2 & 90$\cdot$6 (0$\cdot$42) & 95$\cdot$0 (0$\cdot$15) & 93$\cdot$6 (0$\cdot$17) & 95$\cdot$1 (0$\cdot$13) & 92$\cdot$5 (0$\cdot$14) & 98$\cdot$3 (0$\cdot$06) & 4 & 0$\cdot$3 \\ 
  New L1 & 90$\cdot$7 (0$\cdot$28) & 92$\cdot$9 (0$\cdot$15) & 92$\cdot$4 (0$\cdot$18) & 94$\cdot$9 (0$\cdot$15) & 91$\cdot$2 (0$\cdot$16) & 93$\cdot$5 (0$\cdot$11) & 4 & 3$\cdot$3 \\ 
  New L2 & 92$\cdot$2 (0$\cdot$29) & 93$\cdot$7 (0$\cdot$16) & 93$\cdot$6 (0$\cdot$19) & 96$\cdot$1 (0$\cdot$16) & 94$\cdot$5 (0$\cdot$17) & 95$\cdot$5 (0$\cdot$12) & 4 & 3$\cdot$3 \\ 
  Full RS & 94$\cdot$8 (0$\cdot$59) & 95$\cdot$9 (0$\cdot$21) & 96$\cdot$7 (0$\cdot$24) & 96$\cdot$2 (0$\cdot$21) & 96$\cdot$1 (0$\cdot$22) & 95$\cdot$9 (0$\cdot$21) & 4 & 6 \\ 
  Full WB & 91$\cdot$0 (0$\cdot$54) & 97$\cdot$4 (0$\cdot$18) & 95$\cdot$9 (0$\cdot$22) & 97$\cdot$6 (0$\cdot$18) & 94$\cdot$6 (0$\cdot$20) & 96$\cdot$1 (0$\cdot$19) & 4 & 6 \\ 
  TS AL RS & 94$\cdot$8 (0$\cdot$51) & 96$\cdot$6 (0$\cdot$21) & 96$\cdot$3 (0$\cdot$27) & 97$\cdot$1 (0$\cdot$23) & 95$\cdot$6 (0$\cdot$23) & 98$\cdot$2 (0$\cdot$26) & 4 & 0$\cdot$3 \\ 
  TS AL WB & 91$\cdot$5 (0$\cdot$47) & 95$\cdot$5 (0$\cdot$16) & 94$\cdot$2 (0$\cdot$21) & 96$\cdot$0 (0$\cdot$17) & 92$\cdot$4 (0$\cdot$19) & 97$\cdot$7 (0$\cdot$21) & 4 & 0$\cdot$3 \\ 
  TS L RS & 94$\cdot$1 (0$\cdot$52) & 96$\cdot$2 (0$\cdot$22) & 95$\cdot$6 (0$\cdot$27) & 96$\cdot$0 (0$\cdot$23) & 95$\cdot$4 (0$\cdot$24) & 96$\cdot$3 (0$\cdot$26) & 4 & 3$\cdot$3 \\ 
   TS L WB& 92$\cdot$1 (0$\cdot$49) & 94$\cdot$7 (0$\cdot$18) & 94$\cdot$3 (0$\cdot$22) & 95$\cdot$9 (0$\cdot$19) & 93$\cdot$3 (0$\cdot$20) & 95$\cdot$8 (0$\cdot$21) & 4 & 3$\cdot$3 \\
  Oracle RS & - & 97$\cdot$1 (0$\cdot$21) & 97$\cdot$9 (0$\cdot$26) & 97$\cdot$0 (0$\cdot$20) & 97$\cdot$2 (0$\cdot$18) & - & 4 & 0 \\ 
Oracle WB & - & 97$\cdot$7 (0$\cdot$15) & 95$\cdot$9 (0$\cdot$19) & 98$\cdot$2 (0$\cdot$15) & 97$\cdot$2 (0$\cdot$16) & - & 4 & 0 \\ [15pt]
  & & &  $\tau=$0$\cdot$7 & $n=250$ &  & & & \\[5pt]
  New AL1  & 89$\cdot$6 (0$\cdot$35) & 94$\cdot$8 (0$\cdot$10) & 92$\cdot$2 (0$\cdot$09) & 94$\cdot$9 (0$\cdot$08) & 93$\cdot$6 (0$\cdot$09) & 98$\cdot$7 (0$\cdot$04) & 5 & 0$\cdot$1 \\ 
  New AL2 & 89$\cdot$8 (0$\cdot$34) & 94$\cdot$1 (0$\cdot$09) & 91$\cdot$7 (0$\cdot$09) & 95$\cdot$0 (0$\cdot$08) & 93$\cdot$1 (0$\cdot$09) & 99$\cdot$0 (0$\cdot$04) & 5 & 0$\cdot$1 \\ 
  New L1 & 90$\cdot$1 (0$\cdot$34) & 94$\cdot$4 (0$\cdot$10) & 94$\cdot$2 (0$\cdot$10) & 95$\cdot$4 (0$\cdot$08) & 95$\cdot$1 (0$\cdot$09) & 95$\cdot$4 (0$\cdot$06) & 5 & 2$\cdot$6 \\ 
  New L2 & 90$\cdot$7 (0$\cdot$35) & 94$\cdot$9 (0$\cdot$10) & 94$\cdot$2 (0$\cdot$10) & 95$\cdot$4 (0$\cdot$08) & 95$\cdot$1 (0$\cdot$09) & 95$\cdot$9 (0$\cdot$06) & 5 & 2$\cdot$6 \\ 
  Full RS & 94$\cdot$9 (0$\cdot$39) & 96$\cdot$8 (0$\cdot$12) & 95$\cdot$3 (0$\cdot$12) & 95$\cdot$8 (0$\cdot$10) & 96$\cdot$4 (0$\cdot$11) & 95$\cdot$9 (0$\cdot$11) & 5 & 5 \\ 
    Full WB & 90$\cdot$6 (0$\cdot$37) & 96$\cdot$3 (0$\cdot$11) & 95$\cdot$5 (0$\cdot$11) & 97$\cdot$3 (0$\cdot$09) & 96$\cdot$1 (0$\cdot$11) & 96$\cdot$2 (0$\cdot$10) & 5 & 5 \\ 
 TS AL RS  & 93$\cdot$8 (0$\cdot$37) & 95$\cdot$4 (0$\cdot$12) & 96$\cdot$1 (0$\cdot$10) & 95$\cdot$9 (0$\cdot$11) & 96$\cdot$4 (0$\cdot$12) & 98$\cdot$8 (0$\cdot$11) & 5 & 0$\cdot$1 \\ 
 TS AL WB & 91$\cdot$7 (0$\cdot$35) & 95$\cdot$2 (0$\cdot$11) & 95$\cdot$7 (0$\cdot$09) & 95$\cdot$8 (0$\cdot$10) & 96$\cdot$5 (0$\cdot$11) & 98$\cdot$9 (0$\cdot$11) & 5 & 0$\cdot$1 \\ 
 TS L RS & 93$\cdot$8 (0$\cdot$37) & 95$\cdot$0 (0$\cdot$12) & 95$\cdot$3 (0$\cdot$11) & 96$\cdot$2 (0$\cdot$11) & 95$\cdot$5 (0$\cdot$12) & 96$\cdot$1 (0$\cdot$11) & 5 & 2$\cdot$6 \\ 
  TS L WB & 91$\cdot$2 (0$\cdot$35) & 94$\cdot$8 (0$\cdot$12) & 95$\cdot$2 (0$\cdot$10) & 95$\cdot$7 (0$\cdot$11) & 96$\cdot$8 (0$\cdot$12) & 96$\cdot$0 (0$\cdot$10) & 5 & 2$\cdot$6 \\
  Oracle RS & 94$\cdot$0 (0$\cdot$38) & 96$\cdot$8 (0$\cdot$11) & 95$\cdot$3 (0$\cdot$11) & 95$\cdot$9 (0$\cdot$09) & 96$\cdot$4 (0$\cdot$10) & - & 5 & 0 \\ 
Oracle WB & 90$\cdot$8 (0$\cdot$36) & 95$\cdot$7 (0$\cdot$10) & 94$\cdot$9 (0$\cdot$10) & 96$\cdot$6 (0$\cdot$08) & 96$\cdot$4 (0$\cdot$10) & - & 5 & 0 \\ [5pt]  
\end{tabular}
}
}
\label{tab1}
\begin{tablenotes}
      \footnotesize
\item 
    New AL1: proposed method with adaptive $L_1$ penalty ($\gamma=1$);
   New AL2: proposed method with adaptive $L_1$ penalty  ($\gamma=2$);
 New L1:  proposed method with $L_1$ penalty (data-driven choice of $a_n$);
 New L2:  proposed method with $L_1$ penalty ($a_n=n^{-1/3}$);
 Full RS:  full model with rank-score method; Full WB: full model with wild residual bootstrap;
 TS AL RS: two-step procedure, adaptive $L_1$  ($\gamma=1$) followed by  rank-score
 method;
  TS AL WB: two-step procedure, adaptive $L_1$  ($\gamma=1$) followed by wild residual bootstrap;
 TS L RS: two-step procedure, lasso followed by  rank-score method;
  TS L WB: two-step procedure, lasso followed by wild residual bootstrap;
Oracle RS: oracle model with rank-score method;  Oracle WB: oracle model with wild residual bootstrap;
Zeros: the reported average coverage probability (length) is the average for all zero coefficients;
TP: average number of true positives; FP: average number of false positives.
    \end{tablenotes}
\end{table}

Let
$
  Y = $0$\cdot$25$X_3 + $0$\cdot$5$X_5 + X_7+2X_2+X_1\xi,
$
where $\xi \sim N(0, 1)$ denotes the random error. 
Let $\widetilde{X} = (\widetilde{X}_1,\dots,\widetilde{X}_{10})^T\sim N_{10}(0,I_p)$.
We set $X_1=\Phi(\widetilde{X}_1)$, where $\Phi$ is the standard normal cumulative distribution function, and $X_i=\widetilde{X}_i$ for $i=2,\dots,10$.
We consider estimating the conditional median and the 0$\cdot$7 conditional quantile of $Y$.  Note that the variable $X_1$ is
inactive for estimating the conditional median and is active  for estimating the 0$\cdot$7 conditional quantile.
Let $\vbeta=(\beta_1, \ldots, \beta_{10})^T$ be the vector of quantile regression coefficients. We have 
$\beta_3=$0$\cdot$25, $\beta_5=$0$\cdot$5, $\beta_7=1$, $\beta_9=2$,
$\beta_2=\beta_4=\beta_6=\beta_8=\beta_{10}=0$ for both quantiles, $\beta_1=0$ for the conditional median and
$\beta_1=\Phi^{-1}$(0$\cdot$7) for the 0$\cdot$7 conditional quantile.

We perform 1000 simulations with 400 bootstrapped samples for each.  
We report sample size $n=100$ for estimating the conditional median and size 250 for estimating the 0$\cdot$7 conditional quantile, as it is known to be more challenging to estimate a higher quantile than 
to estimate the median.
Table 1 summarizes the simulation results.
The standard errors of the coverage probabilities are below 0.01 and  
the standard errors of the confidence interval lengths are below 0.005 for all cases. We also report the average 
number of nonzero coefficients correctly identified to be nonzero
and the average number of zero coefficients incorrectly identified to be nonzero.
For the two-step procedure, we only report results for $\gamma=1$ if adaptive lasso
is applied in Step 1 as the results for $\gamma=2$ are similar.
Additional simulation results are given in the Supplementary Material.

The wild residual bootstrap procedures achieve the specified coverage probability.
For the $L_1$ penalty,  the two choices of  $a_n$ yield similar results.
The adaptive $L_1$ penalty produces sparser models than the $L_1$ penalty does.
The  
resulting confidence intervals are generally shorter than those based on the full model or the two-step procedure.
For the adaptive lasso, the coverage probability of the confidence interval for zero coefficients
is close to one, see Remark 4. 
Similar numerical findings for adaptive lasso penalized least square regression 
were reported in Minnier et al. (2011) and Camponovo (2015).

\subsection{A real data example}
We analyze data on the effects of ozone on school children's lung growth \citep{I1}. 
The study was carried out from February 1996 to October 1999 in South
Western Germany on school children initially in first and second primary school classes.
The data we analyze contain a subset of 496 children with complete data at three 
examinations \citep{B2}. 

The response variable is the forced vital capacity
of the lung. We consider the ten explanatory 
variables with the largest
inclusion probabilities using the bootstrap procedure from \citet{DB}:
gender, $x_1$; height at pulmonary function testing, $x_2$; 
weight at pulmonary function testing, $x_3$; 
maximal nitrogen oxide value of last 24 hours before pulmonary function testing, $x_4$;
wheezing or whistling in the chest, $x_5$; shortness of breath, $x_6$;
whether patient lives in a village with high ozone values, $x_7$;
sensitization to pollens, $x_8$; sensitization to dust mite allergens, $x_9$;
and age at March 1, 1996, $x_{10}$.

Table~\ref{ozone} reports 95\% confidence intervals
for each covariate from bootstrapping 
penalized quantile regression with 
the adaptive $L_1$ and $L_1$ penalties
for estimating the conditional median and the conditional 0.7 quantile.  
For both methods, the variables  $x_1$,  $x_2$ and $x_3$ are identified as significant at both quantiles.

\begin{table}[ht]
\caption{Analysis of ozone data: wild residual-based 95\% bootstrapped 
 confidence intervals for the 0.5 and 0.7 conditional quantiles}
 {\scalebox{0.8}{
\centering
\begin{tabular}{lllllllll}
  \\
    & \multicolumn{3}{c}{$\tau=$0$\cdot$5} && 
    & \multicolumn{3}{c}{$\tau=$0$\cdot$7}\\
  & New AL1&  New AL2 & New L && New AL1 &  New AL2 & New L\\ 
Intercept & (2$\cdot$26, 2$\cdot$31) & (2$\cdot$26, 2$\cdot$30) & (2$\cdot$26, 2$\cdot$31) && (2$\cdot$37, 2$\cdot$41) & (2$\cdot$37, 2$\cdot$41) & (2$\cdot$37, 2$\cdot$42)\\ 
  $x_1$ & ($-$0$\cdot$13, $-$0$\cdot$08) & ($-$0$\cdot$12, $-$0$\cdot$09) & ($-$0$\cdot$10, $-$0$\cdot$10)&& ($-$0$\cdot$12, $-$0$\cdot$08) & ($-$0$\cdot$12, $-$0$\cdot$08) & ($-$0$\cdot$10, $-$0$\cdot$10) \\ 
  $x_2$  & (0$\cdot$15, 0$\cdot$22) & (0$\cdot$14, 0$\cdot$20) & (0$\cdot$18, 0$\cdot$24) && (0$\cdot$16, 0$\cdot$22) & (0$\cdot$16, 0$\cdot$22) & (0$\cdot$21, 0$\cdot$26)  \\ 
 $x_3$  & (0$\cdot$04, 0$\cdot$12) & (0$\cdot$05, 0$\cdot$12) & (0$\cdot$07, 0$\cdot$08)  && (0$\cdot$06, 0$\cdot$15) & (0$\cdot$06, 0$\cdot$15) & (0$\cdot$08, 0$\cdot$09)\\ 
  $x_4$  & (0, 0) & ($-$0$\cdot$01, 0$\cdot$01) & (0, 0) && ($-$0$\cdot$01, 0) & ($-$0$\cdot$01, 0) & (0, 0) \\ 
  $x_5$ & (0, 0) & (0$\cdot$01, 0$\cdot$03) & (0$\cdot$02, 0$\cdot$02) && ($-$0$\cdot$01, 0) & ($-$0$\cdot$01, 0) & (0, 0) \\ 
 $x_6$  & (0, 0) & (0, 0) & (0, 0) && (0$\cdot$01, 0$\cdot$05) & (0$\cdot$01, 0$\cdot$05) & (0$\cdot$03, 0$\cdot$03)\\ 
  $x_7$  & (0, 0) & ($-$0$\cdot$01, 0$\cdot$01) & (0, 0) && (0, 0$\cdot$01) & ($-$0$\cdot$01, 0$\cdot$01) & (0, 0) \\ 
 $x_8$  & (0, 0) & ($-$0$\cdot$01, 0$\cdot$01) & (0, 0)  && ($-$0$\cdot$03, $-$0$\cdot$01) & ($-$0$\cdot$03, 0) & ($-$0$\cdot$02, $-$0$\cdot$02)  \\ 
  $x_9$  & (0, 0) & ($-$0$\cdot$01, 0$\cdot$01) & (0, 0) && (0, 0$\cdot$02) & (0, 0$\cdot$02) & (0, 0)\\ 
  $x_{10}$  & (0, 0) & (0, 0$\cdot$04) & (0$\cdot$01, 0$\cdot$02) && (0, 0$\cdot$01) & (0, 0$\cdot$01) & ($-$0$\cdot$01, 0)  \\[5pt]
\end{tabular}}}
\label{ozone}
\begin{tablenotes}
      \footnotesize
\item 
New AL1: proposed method with adaptive $L_1$ penalty ($\gamma=1$);
New AL2: proposed method with adaptive $L_1$ penalty ($\gamma=2$);
  and New L: proposed method with $L_1$ penalty (data-driven choice of $a_n$).
    \end{tablenotes}
\end{table}

\appendix

\section*{Appendix: Proofs of Theorems \ref{Balasso} and \ref{Bqlasso}}
We use $E^*$ and $\mbox{var}^*$ to denote expectation and variance conditional on the 
sample $\vz$.  
Let $E_{\vr, \vz}$ and  $\mbox{var}_{\vr, \vz}$
be the expectation and variance with respect to the joint distribution 
of $\vr$ and $\vz$.
Let $\mbox{pr}$ denote the probability under the joint distribution; and 
let $\mbox{pr}_{\vr\mid \vz}$ denote the probability of $\vr$
conditional on $\vz$. 
A random variable $R_n$ is said to be $o_{p_{\vr}}^*(1)$ if
for any $\epsilon, \delta>0$, 
$\mbox{pr}_{\vz}\{\mbox{pr}_{\vr|\vz}(|R_n|>\epsilon)>\delta\}\ra 0$, as $n\ra \infty$, 
and $o_{\mbox{pr}_{\vr,\vz}}(1)$ is the regular
notion with respect to the joint distribution of $\vr$ and $\vz$.
Lemma 3 from \cite{CH} will be used repeatedly to
allow for the transition of various stochastic orders in different probability spaces.

Let
$
V_n^*(\vdelta)=\sum_{i=1}^n\big\{\rho_{\tau}(\epsilon_i^*-n^{-1/2}\vx_i^T\vdelta)-\rho_{\tau}(\epsilon_i^*)\big\}.
$
Let $\psi_{\tau}(u)=\tau-I(u<0)$. 
It follows from \cite{knight1998} and \cite{K2} that 
\bqa
V_n^*(\vdelta)&=&-n^{-1/2}\sum_{i=1}^n\vx_i^T\vdelta\psi_{\tau}(\epsilon_i^*)
+\sum_{i=1}^n\int_{0}^{n^{-1/2}\vx_i^T\vdelta}
\big\{I(\epsilon_i^*\leq s)-I(\epsilon_i^*\leq 0)\big\}ds\\
&=&V_{1n}^*(\vdelta)+V_{2n}^*(\vdelta).
\eqa

\blem \label{V11} 
Under the conditions of Theorem \ref{Balasso},
\bqan\label{dog1}
\sup_{t}\big| \textnormal{pr}_{\vr\mid \vz}\{V_{1n}^*(\vdelta)\leq t\}-
 \textnormal{pr}_{\vz}\{ -\vdelta^TH\leq t\}\big|=o_{ \textnormal{pr}_{\vz}}(1).
\eqan
\elem
The proof of Lemma A1 is given in the Supplementary Material. 

\blem\label{apple0}
Under the conditions of Theorem \ref{Balasso}, 
\bqan\label{dog2} 
V_{2n}^*(\vdelta)= \vdelta^TB_1\vdelta/2+o_{p_{\vr}}^*(1).
\eqan
\elem
\noindent{{\it Proof.}} Recall  $\epsilon_i^*=r_i|\hat{\epsilon}_i|$ and
$\hat{\epsilon}_i=\epsilon_i-\vx_i^T(\widetilde{\vbeta}-\vbeta_0)$.
We will show that 
\bqa
\sup_{\vb \in B} |V_{2n}^*(\vdelta,\vb) - \vdelta^TB_1\vdelta/2| =o_{p_{\vr}}^*(1), 
\eqa
where $V_{2n}^*(\vdelta,\vb) = \sum_{i=1}^n\int_{0}^{n^{-1/2}\vx_i^T\vdelta}
\big\{I(r_i|\epsilon_i - n^{-1/2+\eta} \vx_i^T \vb |\leq s)-I(r_i\leq 0)\big\}ds, $
with $B$ a compact set and $\eta>0$.  Since $\mbox{pr}_{\vz}\{n^{1/2-\eta}(\widetilde 
\vbeta-\vbeta_0) \in B\} \rightarrow 1$,  the result of the lemma follows.
By Lemma 3 of \cite{CH}, it suffices to show that 
\bqa
\sup_{\vb \in B} |V_{2n}^*(\vdelta,\vb) - 
\vdelta^TB_1\vdelta/2| = o_{p_{\vr,\vz}}(1).
\eqa 

We will use Theorem 2.11.9 in \cite{VdVW}.  
For a fixed $\varepsilon>0$, divide the set $B$ in $O(\varepsilon^{-2p})$ cubes of the form $C_\vk = 
\prod_{j=1}^p [b_{j,k_j-1},b_{j,k_j})$ with $\vk=(k_1,\ldots,k_p)^T$, $k_j=1,\ldots,O(\varepsilon^{-2})$ for
$j=1,\ldots,p$, and 
$b_{j,k_j} - b_{j,k_j-1} \le \varepsilon^2$.  Then, writing $V_{2n}^*(\vdelta,\vb) = \sum_{i=1}^n v_{i\vb}$, 
we will show that 
\begin{eqnarray} \label{brack} 
\sum_{i=1}^n E_{\vr,\vz} \Big(\sup_{\vb,\vb' \in C_\vk}  |v_{i\vb} - v_{i\vb'}|^2 \Big) \le \varepsilon^2. 
\end{eqnarray}
Indeed, for fixed $i$ and for $\vb,\vb' \in C_\vk$,  $|v_{i\vb} - v_{i\vb'}|^2 $ is bounded above by
\begin{eqnarray*}
&&  \Big| \int_0^{n^{-1/2} \vx_i^T\vdelta} \big\{I(r_i|\epsilon_i - n^{-1/2+\eta} \vx_i^T \vb |\leq s)-I(r_i|\epsilon_i - n^{-1/2+\eta} \vx_i^T \vb' |\leq s)\big\}ds \Big|^2 \\
&\le & \ I(\vx_i^T \vdelta>0) n^{-1/2} \vx_i^T \vdelta  \int_0^{n^{-1/2} \vx_i^T\vdelta} \big|I(r_i|\epsilon_i - n^{-1/2+\eta} \vx_i^T \vb |\leq s)-I(r_i|\epsilon_i - n^{-1/2+\eta} \vx_i^T \vb' |\leq s)\big|ds \\
&&  + I(\vx_i^T \vdelta \le 0) n^{-1/2} |\vx_i^T \vdelta|  \int_0^{n^{-1/2} |\vx_i^T\vdelta|} \big|I(r_i|\epsilon_i - n^{-1/2+\eta} \vx_i^T \vb |\leq -s)-I(r_i|\epsilon_i - n^{-1/2+\eta} \vx_i^T \vb' |\leq -s)\big|ds.
\end{eqnarray*}
Let us focus on the first term above, as the second term is similar. The first term equals
\begin{eqnarray*}
&&  I(\vx_i^T \vdelta>0,r_i>0) n^{-1/2} \vx_i^T \vdelta  \int_0^{n^{-1/2} \vx_i^T\vdelta} \big|I(-s/r_i + n^{-1/2+\eta} \vx_i^T \vb \le \epsilon_i \le s/r_i + n^{-1/2+\eta} \vx_i^T \vb) \\
&& -I(-s/r_i + n^{-1/2+\eta} \vx_i^T \vb' \le \epsilon_i \le s/r_i + n^{-1/2+\eta} \vx_i^T \vb')\big|ds \\
&\le &   I(\vx_i^T \vdelta>0,r_i>0) n^{-1/2} \vx_i^T \vdelta  \int_0^{n^{-1/2} \vx_i^T\vdelta} \Big\{\big|I(\epsilon_i \le s/r_i + n^{-1/2+\eta} \vx_i^T \vb)-I(\epsilon_i \le s/r_i + n^{-1/2+\eta} \vx_i^T \vb')\big| \\
&&  + \big|I(\epsilon_i \le -s/r_i + n^{-1/2+\eta} \vx_i^T \vb)-I(\epsilon_i \le -s/r_i + n^{-1/2+\eta} \vx_i^T \vb')\big| \Big\} ds \\
&\le&  I(\vx_i^T \vdelta>0,r_i>0) n^{-1/2} \vx_i^T \vdelta  \int_0^{n^{-1/2} \vx_i^T\vdelta} \Big[\big\{
I(\epsilon_i \le s/r_i + n^{-1/2+\eta} \vx_i^T \vb_\vk)-I(\epsilon_i \le s/r_i + n^{-1/2+\eta} \vx_i^T 
\vb_{\vk-1})\big\} \\
&&  + \big\{I(\epsilon_i \le -s/r_i + n^{-1/2+\eta} \vx_i^T \vb_\vk)-I(\epsilon_i \le -s/r_i + 
n^{-1/2+\eta} \vx_i^T \vb_{\vk-1})\big\} \Big] ds,
\end{eqnarray*}
where for notational simplicity we assume that all components of $\vx_i$ are positive.  Hence, 
\begin{eqnarray*}
&& \sum_{i=1}^n E_{\vr,\vz} \Big(\sup_{\vb,\vb' \in C_\vk}  |v_{i\vb} - v_{i\vb'}|^2 \Big) \\
&\le &  n^{-1/2} \sum_{i=1}^n |\vx_i^T \vdelta| \int  \int_0^{n^{-1/2} |\vx_i^T\vdelta|} \Big[\big\{F_i(s/r + n^{-1/2+\eta} \vx_i^T \vb_\vk)-F_i(s/r + n^{-1/2+\eta} \vx_i^T \vb_{\vk-1})\big\} \\
\hspace*{-3cm} && \hspace*{.5cm} + \big\{F_i(-s/r + n^{-1/2+\eta} \vx_i^T \vb_\vk)-F_i(-s/r + n^{-1/2+\eta} \vx_i^T \vb_{\vk-1})\big\} \Big] ds \, dG(r) \\
&\le &  2 n^{-1} \sum_{i=1}^n |\vx_i^T \vdelta|^2 n^{-1/2+\eta} \vx_i^T |\vb_\vk-\vb_{\vk-1}| \sup_{t \in {\cal N}_i} f_i(t)  \le c \varepsilon^2,
\end{eqnarray*} 
for some $0 < c < \infty$, for $\eta \le 1/2$, where ${\cal N}_i$ is a neighborhood of 0 such that $\sup_{t \in {\cal N}_i} f_i(t) < \infty$; see Condition 1.  This verifies (\ref{brack}). 

Let $N_{[\,]}(\varepsilon,B,L_2^n)$ be the bracketing number of $B$, i.e.,\ the minimal number of 
sets $N_\varepsilon$ in a partition $B = \cup_{j=1}^{N_\varepsilon} B_{\varepsilon j}$ such that 
$\sum_{i=1}^n E_{\vr,\vz} \big\{\sup_{\vb,\vb' \in B_{\varepsilon j}}  (v_{i\vb} - v_{i\vb'})^2 \big\} \le \varepsilon^2$ for 
$j=1,\ldots,N_\varepsilon$.  For any $\delta_n \downarrow 0$, 
$$ 
\int_0^{\delta_n} \{\log N_{[\,]}(\varepsilon,B,L_2^n)\}^{1/2} \, d\varepsilon \le c 
\int_0^{\delta_n} \{\log (\varepsilon^{-2p})\}^{1/2} d\varepsilon \rightarrow 0. 
$$
Since the partition of $B$ does not depend on $n$ and since $\sup_{\vb \in B}|v_{i\vb}| 
\rightarrow 0$ for all $i$, it follows from Theorem 2.11.9 in \cite{VdVW} that 
$V_{2n}^*(\vdelta,\vb) - E_{\vr,\vz}\{V_{2n}^*(\vdelta,\vb)\}$ converges weakly in $\ell^\infty(B)$ provided
it converges marginally, where $\ell^\infty(B)$ is the space of bounded functions from $B$ to 
$\mathcal{R}$ equipped with the supremum norm.    

To check convergence of $V_{2n}^*(\vdelta,\vb)$ for fixed $\vb \in B$, it suffices to show that 
$\mbox{E}_{\vr,\vz}\{V_{2n}^*(\vdelta,\vb)\}\ra  \vdelta^TB_1\vdelta/2$
and
$\mbox{var}_{\vr,\vz}\{V_{2n}^*(\vdelta,\vb)\}\ra  0$.
Note that
\bqa
&&E_{\vr,\vz}\{V_{2n}^*(\vdelta,\vb)\} \\
&=&  E_{\vr}\Big(E_{\vz|\vr}\Big[\sum_{i=1}^n\int_{0}^{n^{-1/2}\vx_i^T\vdelta}
\big\{I(r_i|\epsilon_i - n^{-1/2+\eta} \vx_i^T \vb |\leq s)-I(r_i\leq 0)\big\}ds\Big]\Big)\\
&=&  \int_{0}^{\infty}\sum_{i=1}^n \int_{0}^{n^{-1/2}\vx_i^T\vdelta}
\big\{F_i(s/r+n^{-1/2+\eta} \vx_i^T \vb)-F_i(-s/r+n^{-1/2+\eta} \vx_i^T \vb)\big\}I(\vx_i^T\vdelta>0)ds dG(r) \\
&&  +  \int_{-\infty}^{0}\sum_{i=1}^n \int_{0}^{n^{-1/2}\vx_i^T\vdelta}
\big\{1-F_i(s/r+n^{-1/2+\eta} \vx_i^T \vb)+F_i(-s/r+n^{-1/2+\eta} \vx_i^T \vb)-1\big\}I(\vx_i^T\vdelta<0)ds dG(r)\\
&& = W_1+W_2,
\eqa
say, where $F_i$ denotes the distribution of $\epsilon_i$.
\bqa
W_1&=& \int_{0}^{\infty}\sum_{i=1}^n \int_{0}^{n^{-1/2}\vx_i^T\vdelta}
\big\{f_i(0)2s/r\big\}I(\vx_i^T\vdelta>0)ds dG(r)\\
&&+  \int_{0}^{\infty}\sum_{i=1}^n \int_{0}^{n^{-1/2}\vx_i^T\vdelta}
\big\{f_i(t^*/r)-f_i(0)\big\}2s/rI(\vx_i^T\vdelta>0)ds dG(r)=W_{11}+W_{12},
\eqa
say, where $t^*$ is between $-n^{-1/2}\vx_i^T\vdelta+n^{-1/2+\eta} \vx_i^T \vb $ and $n^{-1/2}\vx_i^T\vdelta+n^{-1/2+\eta} \vx_i^T \vb$. 
Note that
\bqa
W_{11}=\int_{0}^{\infty}r^{-1}dG(r) \sum_{i=1}^nf_i(0)\big(n^{-1/2}\vx_i^T\vdelta\big)^2I(\vx_i^T\vdelta>0)
= \frac{1}{2}\vdelta^T\big\{n^{-1} \sum_{i=1}^nf_i(0)\vx_i\vx_i^TI(\vx_i^T\vdelta>0)\big\}\vdelta.
\eqa
By Condition 1, there exists a positive constant $c$ such that
\bqa
|W_{12}|&\leq &c\int_{0}^{\infty}\sum_{i=1}^n \int_{0}^{n^{-1/2}\vx_i^T\vdelta}
\big(n^{-1/2}\vx_i^T\vdelta/r + n^{-1/2+\eta} |\vx_i^T \vb| \big)2s/rI(\vx_i^T\vdelta>0)ds dG(r)\\
&\leq &c\Big\{\int_{0}^{\infty}r^{-2}dG(r)\Big\} 
\big(n^{-1/2} ||\vdelta||\max_{1\le i \leq n}||\vx_i||\big)
\Big[\vdelta^T\big\{n^{-1} \sum_{i=1}^n \vx_i\vx_i^TI(\vx_i^T\vdelta>0)\big\}\vdelta\Big] \\
&& + c\Big\{\int_{0}^{\infty}r^{-1}dG(r)\Big\} 
\big(n^{-1/2+\eta}||\vb||\max_{1\le i \leq n}||\vx_i|| \big)
\Big[\vdelta^T\big\{n^{-1} \sum_{i=1}^n \vx_i\vx_i^TI(\vx_i^T\vdelta>0)\big\}\vdelta\Big]
\ra 0,
\eqa
as Conditions 3 and 4 imply that $\int_0^{\infty}r^{-2}dG(r)$ is bounded, and by
Condition 2 we have $n^{-1/2+\eta}\max_{1\le i \leq n}||\vx_i||\ra 0$ for $\eta$ small enough. 
Similarly, we can show
$
W_2=\frac{1}{2}\vdelta^T\big\{n^{-1} \sum_{i=1}^nf_i(0)\vx_i\vx_i^TI(\vx_i^T\vdelta<0)\big\}\vdelta+o(1).
$
Hence, $E_{\vr,\vz}\{V_{2n}^*(\vdelta,\vb)\}\ra  \vdelta^TB_1\vdelta/2$ as $n\ra \infty$.
To show $\mbox{var}_{\vr,\vz}\{V_{2n}^*(\vdelta,\vb)\}\ra  0$, 
we have
\bqa
\mbox{var}_{\vr,\vz}\{V_{2n}^*(\vdelta,\vb)\}&=&
\sum_{i=1}^n\mbox{var}_{\vr,\vz}\Big[\int_{0}^{n^{-1/2}\vx_i^T\vdelta}
\big\{I(r_i|\epsilon_i-n^{-1/2+\eta} \vx_i^T \vb|\leq s)-I(r_i\leq 0)\big\}ds\Big]
\\
&\leq & \sum_{i=1}^nE_{\vr,\vz}\Big[\int_{0}^{n^{-1/2}\vx_i^T\vdelta}
\big\{I(r_i |\epsilon_i-n^{-1/2+\eta} \vx_i^T \vb|\leq s)-I(r_i\leq 0)\big\}ds\Big]^2\\
&=& \Big(n^{-1/2}||\vdelta|| \max_{1\le i \leq n}||\vx_i|| \Big)E_{\vr,\vz}\{V_{2n}^*(\vdelta,\vb)\},
\eqa
where the last equality follows because $\int_{0}^{n^{-1/2}\vx_i^T\vdelta}
\big\{I(r_i|\epsilon_i-n^{-1/2+\eta} \vx_i^T \vb|\leq s)-I(r_i\leq 0)\big\}ds$ is always nonnegative.
Since $n^{-1/2}\max_{1\leq n}||\vx_i||\ra 0$ and $\mbox{E}_{\vr,\vz}\{V_{2n}^*(\vdelta,\vb)\}\ra  \vdelta^TB_1\vdelta/2$, 
we have $\mbox{var}_{\vr,\vz}\{V_{2n}^*(\vdelta,\vb)\}\ra 0$ as $n\ra \infty$.  This finishes the proof. $\Box$ \\

\noindent {Proof of Theorem \ref{Balasso}}. 
Recall that $Q_n^*(\vdelta)=\sum_{i=1}^n\big\{\rho_{\tau}(\epsilon_i^*-n^{-1/2}\vx_i^T\vdelta)
-\rho_{\tau}(\epsilon_i^*)\big\}
+\lambda_n\sum_{j=1}^pw_j^*\big(|\widetilde{\beta}_j+n^{1/2}\delta_j|-|\widetilde{\beta}_j|\big)$,
where $w_j^*=|\overline{\beta}_j^*|^{-\gamma}$,
$\overline{\vbeta}^*=(\overline{\beta}^*_0, \overline{\beta}^*_1, \ldots, \overline{\beta}^*_p)^T$
is the ordinary quantile regression estimator computed from the bootstrap sample, $\gamma>0$.
We have
$n^{1/2}(\widetilde{\vbeta}^*-\widetilde{\vbeta})=\argmin_{\vdelta}Q_n^*(\vdelta).
$
Let $A_n$ denote the event that the adaptive lasso estimator
$\widetilde{\vbeta}$ correctly estimated all the zero components of $\vbeta$, i.e., $A_n$ is the set of
all $\omega\in \Omega$ such that
$\{j: 1\leq j\leq p, \widetilde{\beta}_j(\omega)=0\}=\{q+1, \ldots, p\}$.
Then it follows from Lemma \ref{liona} that $\mbox{pr}(A_n)\ra 1$ as $n\ra \infty$.  There exists a subsequence $\{n_k\}$ such that
$\mbox{pr}(A_{n_k}^c i.o.)=0$. Let $\Omega_0^c$ be the union of $\limsup_{k}A_{n_k}^c$
and the event on which (\ref{dog1}) or (\ref{dog2}) fails to hold, then $\mbox{pr}(\Omega_0)=1$.
For any fixed $w\in \Omega_0$, there exists $n_{w}\geq 1$ such that for all 
$n\geq n_{w}$, $\{j: 1\leq j\leq p, \widetilde{\beta}_{nj}(\omega)=0\}=\{q+1, \ldots, p\}$.
Hence on $\Omega_0$, as $n\ra \infty$,
\bqa
Q_n^*(\vdelta)\ra Q^*(\vdelta)=
\begin{cases}
-\vdelta^TH+\vdelta^TB_1\vdelta/2, &\text{$\delta_{q+1}=\cdots=\delta_{p}=0$,}\\
+\infty, &\text{otherwise},
\end{cases}
\eqa
in probability. Following the same argument as in Lemma \ref{liona} and applying epi-convergence theory
see the unpublished technical reports of  Geyer ({\it On the asymptotics of
convex stochastic optimization}, technical report, 1996) and Knight ({\it Epi-convergence in distribution and stochastic
equi-semicontinuity}, technical report, 1999),
the result is established by the equivalent representation of bootstrap consistency in (23.2) of  \citet{Vaart}. $\Box$ \\
\noindent{Proof of Theorem \ref{Bqlasso}.} Let $A_n=\{||\widecheck{\vbeta}^*-\vbeta_0||\leq c n^{-1/2}\log(n)\}$ for some 
given positive constant $c$.
Since $\overline{\vbeta}$ is $n^{1/2}$-consistent, we have $\mbox{pr}(A_n)\ra 1$. 
Let
$
Q_n^{**}(\vdelta)=\sum_{i=1}^n\big\{\rho_{\tau}(\epsilon_i^{**}-n^{-1/2}\vx_i^T\vdelta)-\rho_{\tau}(\epsilon_i^{**})\big\}
+\lambda_n\sum_{j=1}^p\big(|\widecheck{\beta}^*_j+n^{-1/2}\delta_j|-|\widecheck{\beta}^*_j|\big),
$
then $n^{1/2}(\widecheck{\vbeta}^{**}-\widecheck{\vbeta}^{*})$ minimizes $Q_n^{**}(\vdelta)$.
Let
$
V_n^{**}(\vdelta)=\sum_{i=1}^n\big\{\rho_{\tau}(\epsilon_i^{**}-\vx_i^T\vdelta/n^{1/2})-\rho_{\tau}(\epsilon_i^{**})\big\}.
$
We can write
\bqa
V_n^{**}(\vdelta)&=&-n^{-1/2}\sum_{i=1}^n\vx_i^T\vdelta\psi_{\tau}(\epsilon_i^{**})
+\sum_{i=1}^n\int_{0}^{n^{-1/2}\vx_i^T\vdelta}
\big\{I(\epsilon_i^{**}\leq s)-I(\epsilon_i^{**}\leq 0)\big\}ds\\
&=& V_{1n}^{**}(\vdelta)+V_{2n}^{**}(\vdelta).
\eqa
Similarly as in the proof of Lemma A1, 
\bqa
\sup_{t}\big| \textnormal{pr}_{\vr\mid \vz}\{V_{1n}^{**}(\vdelta)\leq t\}-
 \textnormal{pr}_{\vz}\{ -\vdelta^TH\leq t\}\big|=o_{ \textnormal{pr}_{\vz}}(1).
\eqa
Similarly as in the proof of Lemma A2, 
$V_{2n}^{**}(\vdelta)= \vdelta^TB_1\vdelta/2+o_{p_{\vr}}^*(1).$
For $n$ sufficiently large, on the event $A_n$,
$\mbox{sign}(\widecheck{\beta}^*_j)=\mbox{sign}(\beta_{0j})$ and $\widecheck{\beta}^*_j=\overline{\beta}_{0j}$ for $j=1,\ldots, q$;
and $\widecheck{\beta}^*_j=0$ for $j=q+1,\ldots, p$. 
Conditional on the data,
$\lambda_n\sum_{j=1}^p\big\{|\widecheck{\beta}^*_j+\delta_j/n^{1/2}|-|\widecheck{\beta}^*_j|\big\}
\ra  \lambda_0\sum_{j=1}^p\big\{|\delta_j|I(\widecheck{\beta}^*_j=0)
+\delta_j\mbox{sign}(\beta_{0j})I(\widecheck{\beta}^*_j\neq 0)\}$.
For any $1\leq j \leq p$,
\bqa
&& \mbox{pr}\Big\{|\delta_j|I(\widecheck{\beta}^*_j=0)+\delta_j\mbox{sign}(\beta_{0j})I(\widecheck{\beta}^*_j\neq 0)=|\delta_j|I(\beta_{0j}=0)+\delta_j\mbox{sign}(\beta_{0j})I(\beta_{0j}\neq 0)\Big\}\\
&\geq & \mbox{pr}\Big\{|\delta_j|I(\widecheck{\beta}^*_j=0)+\delta_j\mbox{sign}(\beta_{0j})I(\widecheck{\beta}^*_j\neq 0)=|\delta_j|I(\beta_{0j}=0)+\delta_j\mbox{sign}(\beta_{0j})I(\beta_{0j}\neq 0), A_n\Big\}
\ra 1,
\eqa
as $n\ra \infty$.  Therefore, conditional on the data,  as $n\ra \infty$,
\bqa
Q_n^{**}(\vdelta)\ra -\vdelta^TH+\vdelta^TB_1\vdelta/2
+\lambda_0\sum_{j=1}^p\big\{|\delta_j|I(\beta_{0j}=0)+\delta_j\mbox{sign}(\beta_{0j})I(\beta_{0j}\neq 0)\big\},
\eqa
in distribution.
Following the same argument as in Lemma \ref{bird} and applying epi-convergence theory,
see the unpublished technical reports of  Geyer ({\it On the asymptotics of
convex stochastic optimization}, technical report, 1996) and Knight ({\it Epi-convergence in distribution and stochastic
equi-semicontinuity}, technical report, 1999),
the result is established by the equivalent representation of bootstrap consistency in (23.2) of  \citet{Vaart}.
$\Box$

\bibliographystyle{apalike}
\bibliography{WildQR}

\section*{Supplementary Material}

\section*{Appendix 1}
\subsection*{Proofs of Lemma 1, Lemma 2 and Lemma A1}
The proofs of Lemmas 1 and 2 combine the ideas in
\cite{W2} and \cite{wang2012}.
Section 3.3 of \cite{W2} considered an extension of the asymptotic theory
of penalized quantile regression to the general heteroscedastic error setting
but only a sketch of the derivation was provided in their online supplement. We provide a
detailed derivation below for completeness. \\

\noindent{{\it Proof of Lemma 1.}} Write  
$\vdelta=(\vdelta_1^T, \vdelta_2^T)^T$,
where $\vdelta_1=(\delta_0, \delta_1, \ldots, \delta_q)^T$
and $\vdelta_2=(\delta_{q+1}, \ldots, \delta_p)^T$.
Write $\widetilde{\vdelta}=(\widetilde{\vdelta}_1^T, \widetilde{\vdelta}_2^T)^T=n^{1/2}(\widetilde{\vbeta}-\vbeta_0)$.
Then  $\widetilde{\vdelta}$ minimizes $Q_n(\vdelta)$, where
\bqa
Q_n(\vdelta)=\sum_{i=1}^n\big\{\rho_{\tau}(\epsilon_i-n^{-1/2}\vx_i^T\vdelta)-\rho_{\tau}(\epsilon_i)\big\}
+\lambda_n\sum_{j=1}^pw_j\big(|\beta_{0j}+n^{-1/2}\delta_j|-|\beta_{0j}|\big).
\eqa

It follows from \cite{knight1998} and \cite{K2} that
$\sum_{i=1}^n\big\{\rho_{\tau}(\epsilon_i-n^{-1/2}\vx_i^T\vdelta)-\rho_{\tau}(\epsilon_i)\big\}
=-\vdelta^TH+\vdelta^TB_1\vdelta/2+o_p(1),
${}
where $H\sim N\{\vnull, \tau(1-\tau)B_0\}$. For the penalty term, we consider two cases.
(i)
For $j=1,\ldots, q$, $\overline{\beta}_j\ra \beta_{0j}\neq 0$ in probability,
and $n^{1/2}\big(|\beta_{0j}+\delta_j/n^{1/2}|-|\beta_{0j}|\big)\ra \delta_j\mbox{sign}(\beta_{0j})$.
It follows that
$\lambda_nw_j\big(|\beta_{0j}+\delta_j/n^{1/2}|-|\beta_{0j}|\big)=
(n^{-1/2}\lambda_n)|\overline{\beta}_{0j}|^{-\gamma}n^{1/2}\big(|\beta_{0j}+\delta_j/n^{1/2}|-|\beta_{0j}|\big)\ra 0$
as $n^{-1/2}\lambda_n\ra 0$.
(ii) For $j=q+1,\ldots, p$, $\lambda_nw_j\big(|\beta_{0j}+\delta_j/n^{1/2}|-|\beta_{0j}|\big)
=(n^{(\gamma-1)/2}\lambda_n)(n^{1/2}|\overline{\beta}_j|)^{-\gamma}|\delta_j|$.
Since $n^{(\gamma-1)/2}\lambda_n\ra \infty$ and $n^{1/2}|\overline{\beta}_j|=O_p(1)$,
the limit of $\lambda_nw_j\big(|\beta_{0j}+\delta_j/n^{1/2}|-|\beta_{0j}|\big)$ is
zero if $\delta_j=0$ and is $\infty$ if $\delta_j\neq 0$. Hence
\bqa
Q_n(\vdelta)\ra Q(\vdelta)=
\begin{cases}
-\vdelta^TH+\vdelta^TB_1\vdelta/2, &  \delta_{q+1}=\cdots=\delta_{p}=0,\\
+\infty, &\text{otherwise},
\end{cases}
\eqa
in probability. Note that $Q_n(\vdelta)$ is convex in $\vdelta$ and 
its limit $Q(\vdelta)$ has a unique minimum
$(D_1^{-1}W, \vnull_{p-q}^T)^T$, where $W\sim N\{0, \tau(1-\tau)D_0\}$,
$D_0=\lim_{n\ra \infty}n^{-1}\sum_{i=1}^n\vx_{iA}\vx_{iA}^T$
and $D_1=\lim_{n\ra \infty}n^{-1}\sum_{i=1}^nf_i(0)\vx_{iA}\vx_{iA}^T$.
It follows from the epi-convergence theory,
see Geyer ({\it On the asymptotics of
convex stochastic optimization}, technical report, 1996) and Knight ({\it Epi-convergence in distribution and stochastic
equi-semicontinuity}, technical report, 1999),
 that $\widetilde{\vdelta}\ra \argmin_{\vdelta}Q(\vdelta)$ in distribution. Hence
$
\widetilde{\vdelta}_1\ra D_1^{-1}W\sim N\{0, \tau(1-\tau)D_1^{-1}D_0D_1^{-1}\}
$
in distribution and
$\widetilde{\vdelta}_2\ra 0$ in distribution. This proves (ii).

Note that the above asymptotic normality result suggests that 
$\mbox{pr}(j\in \widetilde{A})\ra 1$ for $j=1, \ldots, q$. To prove (i),  it remains to show 
$\mbox{pr}(j\in \widetilde{A})\ra 0$ for $j=q+1, \ldots, p$. 
For a given $j\in \{q+1, \ldots, p\}$, let
\bqa
\xi_j(\vdelta)&=&-\tau n^{-1/2}\sum_{i=1}^nx_{ij}I(\epsilon_i-n^{-1/2}\vx_i^T\vdelta>0)\\
&&+(1-\tau) n^{-1/2}\sum_{i=1}^n x_{ij}I(\epsilon_i-n^{-1/2}\vx_i^T\vdelta<0)-n^{-1/2}\sum_{i=1}^n
x_{ij}v_i+\lambda_n n^{-1/2}w_j\mbox{sign}(\delta_j), 
\eqa where $v_i=0$ if $\epsilon_i-n^{-1/2}\vx_i^T\vdelta\neq 0$
and $v_i\in [\tau-1,\tau]$ otherwise. By the KKT optimality conditions \citep{boyd2004}, if 
$j\in \widetilde{A}$, then there must exist some $v_i^*$ such that
$v_i^*=0$ if $\epsilon_i-\vx_i^T\widetilde{\vdelta}/n^{1/2}\neq 0$
and $v_i^*\in [\tau-1,\tau]$ otherwise, such that
for $\xi_j(\widetilde{\vdelta})$ with $v_i=v_i^*$, $\xi_j(\widetilde{\vdelta})=0$. Hence
$\mbox{pr}(j\in \widetilde{A}) \leq \mbox{pr}\{\xi_j(\widetilde{\vdelta})=0\}$.
Note that
$\lambda_nn^{-1/2}w_j
=(n^{(\gamma-1)/2}\lambda_n)(n^{1/2}|\overline{\beta}_j|)^{-\gamma}\ra \infty$
as $n\ra \infty$.
Furthermore, we have
\bqa
&& -\tau n^{-1/2}\sum_{i=1}^nx_{ij}I(\epsilon_i-n^{-1/2}\vx_i^T\widetilde{\vdelta}>0)
+(1-\tau) n^{-1/2}\sum_{i=1}^n x_{ij}I(\epsilon_i-n^{-1/2}\vx_i^T\widetilde{\vdelta}<0)-n^{-1/2}\sum_{i=1}^n
x_{ij}v_i^*\\
&& = n^{-1/2}\sum_{i=1}^nx_{ij}\{I(\epsilon_i-n^{-1/2}\vx_i^T\widetilde{\vdelta}\leq
0)-\tau\}-n^{-1/2}\sum_{i\in \mathcal{D}}x_{ij}\{v_i^*+(1-\tau)\}, \eqa
where $\mathcal{D}=\{i: \epsilon_i-n^{-1/2}\vx_i^T\widetilde{\vdelta}=0\}$.  
With probability one
the number of elements in $\mathcal{D}$ is finite, following the same argument as  in Section 2.2 of \cite{K2}. 
Therefore, $n^{-1/2}\sum_{i\in \mathcal{D}}x_{ij}\{v_i^*+(1-\tau)\}=O_p(n^{-1/2})$.
Similarly as in the proof of Lemma 4.3 of  \cite{wang2012},
we can show that for any $\Delta>0$, as $n\ra \infty$,
\bqa
&&\sup_{||\vdelta'-\vdelta||\leq
\Delta} n^{-1/2}\Big|\sum_{i=1}^nx_{ij}\big\{I(\epsilon_i-n^{-1/2}\vx_i^T\vdelta'\leq 0)-
I(\epsilon_i-n^{-1/2}\vx_i^T\vdelta\leq 0)\nonumber\\
&&-\mbox{pr}(\epsilon_i-n^{-1/2}\vx_i^T\vdelta'\leq 0)+ \mbox{pr}(\epsilon_i-n^{-1/2}\vx_i^T\vdelta\leq 0)\big\}\Big|=o_p(1),
\eqa
where $||\cdot||$ denotes the $L_2$\bfblue{-}norm.
As a result,
\bqa
&& n^{-1/2}\Big|\sum_{i=1}^nx_{ij}\{I(\epsilon_i-n^{-1/2}\vx_i^T\widetilde{\vdelta}\leq 0)-\tau\}\Big|\\
&\leq & n^{-1/2}\sup_{||\vdelta'-\vdelta||\leq
\Delta}\Big|\sum_{i=1}^nx_{ij}\{I(\epsilon_i-n^{-1/2}\vx_i^T\vdelta'\leq 0)-I(\epsilon_i-n^{-1/2}\vx_i^T\vdelta\leq 0)
-\mbox{pr}(\epsilon_i-n^{-1/2}\vx_i^T\vdelta'\leq 0)\\
&&+ \mbox{pr}(\epsilon_i-n^{-1/2}\vx_i^T\vdelta\leq 0)\}\Big|
+ n^{-1/2}\sup_{||\vdelta'-\vdelta||\leq
\Delta}\Big| \sum_{i=1}^nx_{ij}\{ \mbox{pr}(\epsilon_i-n^{-1/2}\vx_i^T\vdelta'\leq 0)
- \mbox{pr}(\epsilon_i-n^{-1/2}\vx_i^T\vdelta\leq 0)\}\Big|\\
&&+n^{-1/2}\Big|\sum_{i=1}^nx_{ij}\{I(\epsilon_i-n^{-1/2}\vx_i^T\vdelta\leq 0)-\tau\}\Big|\\
&=& o_p(1).
\eqa
Therefore, $ \mbox{pr}(j\in \widetilde{A}) \leq  \mbox{pr}\{\xi_j(\widetilde{\vdelta})=0\}\ra 0$, for $j=q+1, \ldots, p$. $\Box$\\

\noindent{\it Proof of Lemma 2.} Similarly as in the proof of Lemma 1, we can show that
\bqa
&&\sum_{i=1}^n\big\{\rho_{\tau}(\epsilon_i-\vx_i^T\vdelta/n^{1/2})-\rho_{\tau}(\epsilon_i)\big\}
+\lambda_n\sum_{j=1}^p\big(|\beta_{0j}+\delta_j/n^{1/2}|-|\beta_{0j}|\big)\\
&\ra & -\vdelta^TH+\vdelta^TB_1\vdelta/2
+\lambda_0\sum_{j=1}^p\{|\delta_j|I(\beta_{0j}=0)+\delta_j\mbox{sign}(\beta_{0j})I(\beta_{0j}\neq 0)\}
\eqa
in distribution. The result then follows from epi-convergence theory. $\Box$

\noindent{\it Proof of Lemma A1.}  We have
$
V_{1n}^*(\vdelta)= 
n^{-1/2}\sum_{i=1}^n\vx_i^T\vdelta\big\{I(r_i|\hat{\epsilon}_i|<0)-\tau\big\}
=-n^{-1/2}\sum_{i=1}^n\vx_i^T\vdelta\big\{\tau-I(r_i<0)\big\}.
$
Note that $E^*\{V_{1n}^*(\vdelta)\}=0$ and 
$\mbox{var}^*\{V_{1n}^*(\vdelta)\}= \tau(1-\tau)n^{-1}\sum_{i=1}^n\vdelta^T\vx_i\vx_i^T\vdelta\ra  \tau(1-\tau)\vdelta^TB_0\vdelta$
in probability. To check the Lindeberg condition, it suffices to show that $\forall \epsilon>0$,
\bqa
n^{-1}\sum_{i=1}^n\mbox{E}^*\Big(\Big[\vx_i^T\vdelta\big\{I(r_i|\hat{\epsilon}_i|<0)-\tau\big\}\Big]^2
I\big[|\vx_i^T\vdelta\big\{I(r_i|\hat{\epsilon}_i|<0)-\tau\big\}\big|>\epsilon\surd{n}\big]\Big)\ra 0,
\eqa
in probability. This holds by noting that the left side of the above expression is upper bounded
by $n^{-1}\sum_{i=1}^n(\vx_i^T\vdelta)^2
I(|\vx_i^T\vdelta|>\epsilon\surd{n})$, which converges to zero in probability by the dominated convergence theorem.
The result of the lemma follows from the Lindeberg central limit theorem. $\Box$

\section*{Appendix 2}
\subsection*{A Useful Lemma from \cite{CH}}

We use $\vr=\{r_1, \ldots, r_n\}$ to denote the random bootstrap weights and $\vz=\{\vz_1, \ldots, \vz_n\}$ to denote the random sample.
Note that $\vr$ and $\vz$ induce two different sources of randomness.
By the wild bootstrap mechanism, the distribution of $\vr$ is independent of that of $\vz$.
We adopt the following notation from \cite{CH}.
A random quantity $R_n$ is said to be $o_{p_{\vr}}^*(1)$ if
for any $\epsilon, \delta>0$, 
$ \mbox{pr}_{\vz}( \mbox{pr}_{\vr|\vz}(|R_n|>\epsilon)>\delta)\ra 0$, as $n\ra \infty$. Similarly, $R_n$ is said to be $O_{p_{\vr}}^*(1)$ in 
if for all $\delta>0$ there exists a $0<M<\infty$ such that 
$ \mbox{pr}_{\vz}( \mbox{pr}_{\vr|\vz}(|R_n|>M)>\delta)\ra 0$, as $n\ra \infty$. 
And $o_{p_{\vr,\vz}}(1)$, $O_{p_{\vr,\vz}}(1)$ are the regular 
notion with respect to the joint probability distribution of $\vr$ and $\vz$.

The following lemma from \cite{CH} will be used repeatedly in our proof. 
It allows the transition of various stochastic orders in different probability spaces and leads to simplified proofs in many places.

\blem \label{CH2010}
(Lemma 3 of \cite{CH}) Suppose that
\bqa
Q_n=o_{p_{\vr}}^*(1), \quad
R_n=O_{p_{\vr}}^*(1).
\eqa
We have
\bqa
A_n=o_{p_{\vr,\vz}}(1) &\Longleftrightarrow& A_n=o_{p_{\vr}}^*(1), \\
B_n=O_{p_{\vr,\vz}}(1) &\Longleftrightarrow& B_n=O_{p_{\vr}}^*(1),  \\
C_n=Q_n\times O_{p_{\vz}}(1) &\Longleftrightarrow& C_n=o_{p_{\vr}}^*(1), \\
D_n=R_n\times O_{p_{\vz}}(1) &\Longleftrightarrow& D_n=O_{p_{\vr}}^*(1), \\
F_n=Q_n\times R_n &\Longleftrightarrow& F_n=o_{p_{\vr}}^*(1).
\eqa
\elem

\vspace{0.5cm}

\section*{Appendix 3}
\subsection*{Additional Examples of Random Weight Distribution}
The random weights used in the wild residual bootstrap procedure are generated from a distribution $G$ that satisfies 
 Conditions 3--5 of the main paper. Two examples of such random weight distributions were given in \citet{F1}.
We propose below three new weight distributions satisfying these conditions.
Note that compared with the continuous distribution in \citet{F1}, the new distributions given in
Examples 1--2 below have no restrictions on the value of $\tau$.
\\

{\it Example 1.} 
\begin{eqnarray*}
  g_1(r) = G_1'(r) &=& -\frac{r}{8v_1} \mbox{I}\left\{ -2(\tau+v_1) \leq r \leq -2(\tau-v_1) \right\} \\
  && + \frac{r}{8v_2} \mbox{I}\left\{ 2(1-\tau-v_2) \leq r \leq 2(1-\tau+v_2) \right\},
\end{eqnarray*}
where $0<v_1<\tau$ and $0<v_2<1-\tau$.\\

{\it Example 2.} 
\begin{eqnarray*}
  g_2(r) = G_2'(r) &=& -\frac{r}{32v_1} \mbox{I}\left\{ -4(a+v_1) < r < -4(a-v_1) \right\} \\
  && - \frac{r}{32v_2} \mbox{I}\left\{ -4(\tau-a+v_2) < r < -4(\tau-a-v_2) \right\} \\
  && + \frac{r}{32v_3} \mbox{I}\left\{ 4(b-v_3) < r < 4(b+v_2) \right\} \\
  && + \frac{r}{32v_4} \mbox{I}\left\{ 4(1-\tau-b-v_3) < r < 4(1-\tau-b+v_2) \right\},
\end{eqnarray*}
where $0<v_1<a$, $0<v_2<\tau-a$, $0<v_3<b$, $0<v_4<1-\tau-b$, $0<a<\tau$, and $0<b<1-\tau$.\\

{\it Example 3.} 
The point mass distribution
\begin{eqnarray*}
  P(W=r) &=& a\mbox{I}\left\{r=-4a\right\} + (\tau-a)\mbox{I}\left\{r=-4(\tau-a)\right\} \\
  && + b\mbox{I}\left\{r=4b\right\} + (1-\tau-b)\mbox{I}\left\{r=4(1-\tau-b)\right\},
\end{eqnarray*}
where $0<a<\tau$ and $0<b<1-\tau$.

\section*{Appendix 4}
\subsection*{Additional Numerical Results}

In Table \ref{tab:SuppSims}, we summarize the simulation results on the 
comparison of empirical coverage probabilities ($\times100$)  and average interval lengths (in parentheses) for 95\% 
confidence intervals
for $\tau=$0$\cdot$5, $n=250$ and $\tau=$0$\cdot$7, $n=400$ for the various methods described in Section 4.1 of the main paper. 
We note that the standard errors of the coverage probabilities are below 0$\cdot$01 and 
the standard errors of the confidence interval lengths are below 0$\cdot$005 for all cases. 
These results supplement those in Table 1 of the main paper and 
demonstrate further improvement with increased sample size. 

Figure 1 displays the QQ plots of the quantiles of the wild residual bootstrapped estimator versus the empirical quantiles of the
corresponding penalized estimator for estimating the smallest coefficient $\beta_3=$0$\cdot$25 for both the $L_1$ penalty
and the adaptive $L_1$ penalty when sample size $n=250$ and $400$, for $\tau=$0$\cdot$5 and 0$\cdot$7, respectively. Overall, the wild residual bootstrapped distribution 
has satisfactory performance.

\begin{table}
\caption{Empirical coverage probabilities ($\times100$)  and average interval lengths (in parentheses) for nominal 95\% confidence intervals}
{\scalebox{0.9}{
\centering
\begin{tabular}{lcccccccc}
   \\
    & $\beta_1 = \Phi^{-1}(\tau)$ & $\beta_3 =$ 0$\cdot$25 & $\beta_5 =$ 0$\cdot$5 & $\beta_7 = 1$ & $\beta_9 = 2$ & Zeros & TP & FP \\[5pt]
  & & &  $\tau=$0$\cdot$5 & $n=250$ && & & \\[5pt]
 New AL1 & 92$\cdot$4 (0$\cdot$22) & 94$\cdot$8 (0$\cdot$09) & 93$\cdot$3 (0$\cdot$09) & 94$\cdot$9 (0$\cdot$07) & 94$\cdot$0 (0$\cdot$08) & 99$\cdot$0 (0$\cdot$03) & 4 & 0$\cdot$2 \\ 
 New AL2 & 91$\cdot$4 (0$\cdot$28) & 94$\cdot$3 (0$\cdot$09) & 93$\cdot$5 (0$\cdot$09) & 93$\cdot$9 (0$\cdot$07) & 93$\cdot$8 (0$\cdot$08) & 99$\cdot$2 (0$\cdot$03) & 4 & 0$\cdot$2 \\ 
  New L1& 93$\cdot$6 (0$\cdot$14) & 93$\cdot$9 (0$\cdot$10) & 93$\cdot$0 (0$\cdot$09) & 95$\cdot$3 (0$\cdot$08) & 94$\cdot$3 (0$\cdot$09) & 95$\cdot$1 (0$\cdot$05) & 4 & 2$\cdot$8 \\ 
  New L2 & 92$\cdot$6 (0$\cdot$15) & 94$\cdot$4 (0$\cdot$10) & 92$\cdot$9 (0$\cdot$09) & 95$\cdot$4 (0$\cdot$08) & 94$\cdot$5 (0$\cdot$09) & 95$\cdot$0 (0$\cdot$05) & 4 & 2$\cdot$8 \\
Full RS & 93$\cdot$9 (0$\cdot$37) & 97$\cdot$0 (0$\cdot$12) & 95$\cdot$2 (0$\cdot$11) & 96$\cdot$5 (0$\cdot$09) & 96$\cdot$1 (0$\cdot$11) & 96$\cdot$7 (0$\cdot$10) & 4 & 6 \\ 
Full WB & 91$\cdot$5 (0$\cdot$35) & 97$\cdot$0 (0$\cdot$11) & 96$\cdot$2 (0$\cdot$11) & 97$\cdot$2 (0$\cdot$09) & 96$\cdot$2 (0$\cdot$10) & 96$\cdot$7 (0$\cdot$10) & 4 & 6 \\ 
  TS AL RS & 95$\cdot$1 (0$\cdot$34) & 97$\cdot$6 (0$\cdot$14) & 97$\cdot$5 (0$\cdot$11) & 97$\cdot$1 (0$\cdot$13) & 98$\cdot$4 (0$\cdot$15) & 99$\cdot$3 (0$\cdot$11) & 4 & 0$\cdot$2 \\ 
TS AL WB & 93$\cdot$2 (0$\cdot$32) & 93$\cdot$6 (0$\cdot$11) & 95$\cdot$8 (0$\cdot$09) & 95$\cdot$4 (0$\cdot$10) & 97$\cdot$2 (0$\cdot$10) & 99$\cdot$1 (0$\cdot$09) & 4 & 0$\cdot$2 \\ 
  TS L RS & 94$\cdot$7 (0$\cdot$33) & 96$\cdot$7 (0$\cdot$13) & 96$\cdot$8 (0$\cdot$12) & 97$\cdot$5 (0$\cdot$12) & 97$\cdot$8 (0$\cdot$14) & 96$\cdot$9 (0$\cdot$12) & 4 & 2$\cdot$8 \\ 
   TS L WB & 93$\cdot$2 (0$\cdot$32) & 93$\cdot$8 (0$\cdot$11) & 94$\cdot$8 (0$\cdot$10) & 96$\cdot$0 (0$\cdot$10) & 96$\cdot$3 (0$\cdot$11) & 96$\cdot$6 (0$\cdot$10) & 4 & 2$\cdot$8\\
  Oracle RS & - & 98$\cdot$6 (0$\cdot$14) & 96$\cdot$2 (0$\cdot$12) & 98$\cdot$4 (0$\cdot$11) & 97$\cdot$9 (0$\cdot$13) & - & 4 & 0 \\ 
 Oracle WB & - & 96$\cdot$4 (0$\cdot$10) & 95$\cdot$0 (0$\cdot$10) & 96$\cdot$1 (0$\cdot$08) & 95$\cdot$7 (0$\cdot$09) & - & 4 & 0 \\ [15pt]   
  & & &  $\tau=$0$\cdot$7 & $n=400$ && & & \\[5pt]
 New AL1 & 91$\cdot$6 (0$\cdot$27) & 93$\cdot$6 (0$\cdot$07) & 94$\cdot$7 (0$\cdot$06) & 93$\cdot$5 (0$\cdot$07) & 93$\cdot$8 (0$\cdot$07) & 98$\cdot$9 (0$\cdot$03) & 5 & 0$\cdot$1 \\ 
 New AL2 & 91$\cdot$6 (0$\cdot$27) & 93$\cdot$8 (0$\cdot$07) & 94$\cdot$8 (0$\cdot$06) & 93$\cdot$6 (0$\cdot$07) & 94$\cdot$3 (0$\cdot$07) & 99$\cdot$2 (0$\cdot$04) & 5 & 0$\cdot$0 \\ 
 New L1 & 92$\cdot$5 (0$\cdot$27) & 93$\cdot$8 (0$\cdot$08) & 95$\cdot$2 (0$\cdot$07) & 94$\cdot$7 (0$\cdot$08) & 93$\cdot$9 (0$\cdot$07) & 95$\cdot$6 (0$\cdot$04) & 5 & 2$\cdot$1 \\ 
  New L2 & 92$\cdot$8 (0$\cdot$27) & 94$\cdot$0 (0$\cdot$08) & 95$\cdot$6 (0$\cdot$07) & 94$\cdot$6 (0$\cdot$08) & 94$\cdot$0 (0$\cdot$07) & 96$\cdot$1 (0$\cdot$04) & 5 & 2$\cdot$1 \\
   Full RS & 95$\cdot$6 (0$\cdot$30) & 96$\cdot$1 (0$\cdot$09) & 95$\cdot$1 (0$\cdot$08) & 95$\cdot$8 (0$\cdot$09) & 96$\cdot$1 (0$\cdot$08) & 96$\cdot$0 (0$\cdot$08) & 5 & 5 \\ 
Full WB & 93$\cdot$1 (0$\cdot$29) & 95$\cdot$6 (0$\cdot$09) & 96$\cdot$1 (0$\cdot$08) & 95$\cdot$2 (0$\cdot$08) & 95$\cdot$4 (0$\cdot$08) & 95$\cdot$9 (0$\cdot$08) & 5 & 5 \\ 
  TS AL RS & 95$\cdot$3 (0$\cdot$30) & 96$\cdot$3 (0$\cdot$07) & 96$\cdot$1 (0$\cdot$08) & 95$\cdot$4 (0$\cdot$08) & 94$\cdot$5 (0$\cdot$08) & 99$\cdot$3 (0$\cdot$09) & 5 & 0$\cdot$1 \\ 
 TS AL WB & 92$\cdot$7 (0$\cdot$28) & 96$\cdot$6 (0$\cdot$07) & 96$\cdot$9 (0$\cdot$07) & 96$\cdot$1 (0$\cdot$08) & 95$\cdot$5 (0$\cdot$08) & 99$\cdot$2 (0$\cdot$08) & 5 & 0$\cdot$1 \\ 
  TS L RS & 94$\cdot$8 (0$\cdot$30) & 96$\cdot$0 (0$\cdot$08) & 95$\cdot$4 (0$\cdot$08) & 95$\cdot$5 (0$\cdot$08) & 95$\cdot$0 (0$\cdot$08) & 95$\cdot$5 (0$\cdot$08) & 5 & 2$\cdot$1 \\ 
 TS L WB & 92$\cdot$4 (0$\cdot$28) & 96$\cdot$4 (0$\cdot$07) & 96$\cdot$7 (0$\cdot$08) & 95$\cdot$2 (0$\cdot$08) & 95$\cdot$6 (0$\cdot$08) & 96$\cdot$0 (0$\cdot$08) & 5 & 2$\cdot$1 \\
  Oracle RS & 95$\cdot$6 (0$\cdot$30) & 96$\cdot$3 (0$\cdot$08) & 95$\cdot$6 (0$\cdot$07) & 95$\cdot$9 (0$\cdot$08) & 95$\cdot$9 (0$\cdot$08) & - & 5 & 0 \\ 
  Oracle WB & 92$\cdot$8 (0$\cdot$28) & 95$\cdot$8 (0$\cdot$08) & 96$\cdot$8 (0$\cdot$07) & 95$\cdot$7 (0$\cdot$08) & 95$\cdot$4 (0$\cdot$07) & - & 5 & 0 \\ 
[5pt]   
\end{tabular}}}
\begin{tablenotes}
      \footnotesize
\item 
    New AL1: adaptive $L_1$ method with wild residual bootstrap ($\gamma=1$);
   New AL2: adaptive $L_1$ method with wild residual bootstrap  ($\gamma=2$);
 New L1:   $L_1$ method with modified wild residual bootstrap (data-driven choice of $a_n$);
 New L2:   $L_1$ method with modified wild residual bootstrap ($a_n=n^{-1/3}$);
 Full RS:  full model with rank-score method; Full WB: full model with wild residual bootstrap;
 TS AL RS: two-step procedure, adaptive $L_1$  ($\gamma=1$) followed by  rank-score
 method for the refitted model;
  TS AL WB: two-step procedure, adaptive $L_1$  ($\gamma=1$) followed by wild residual bootstrap
for the refitted model;
 TS L RS: two-step procedure, lasso followed by  rank-score method for the refitted model;
  TS L WB: two-step procedure, lasso followed by wild residual bootstrap for the refitted model;
Oracle RS: oracle model with rank-score method;  Oracle WB: oracle model with wild residual bootstrap;
Zeros: the reported average coverage probability (length) is the average for all zero coefficients;
TP: average number of true positives; FP: average number of false positives.
    \end{tablenotes}
\label{tab:SuppSims}
\end{table}

\begin{figure}
\begin{tabular}{ll}
~~~~~~~(a) & ~~~~~~~(b) \\
  \includegraphics[width=65mm]{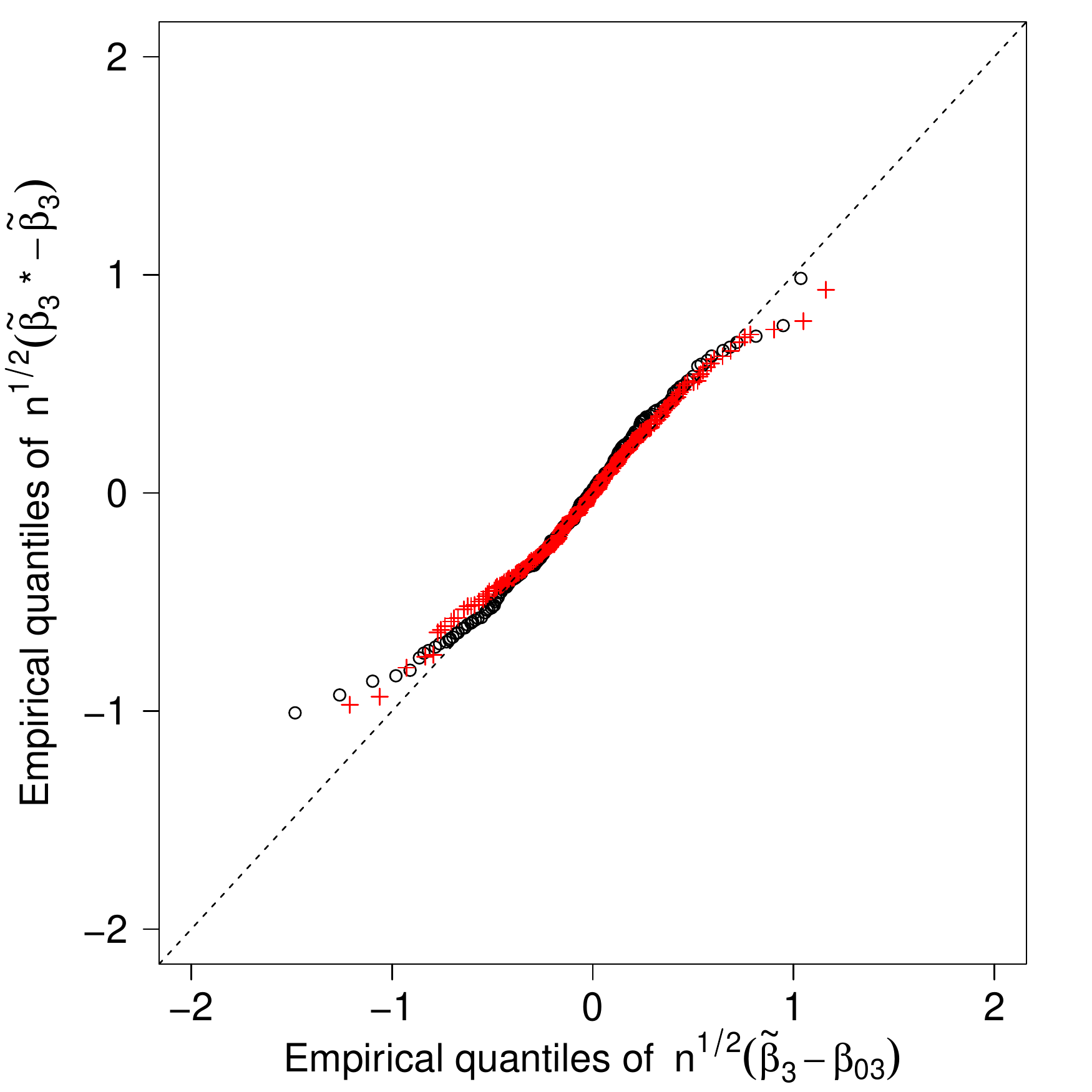} &   \includegraphics[width=65mm]{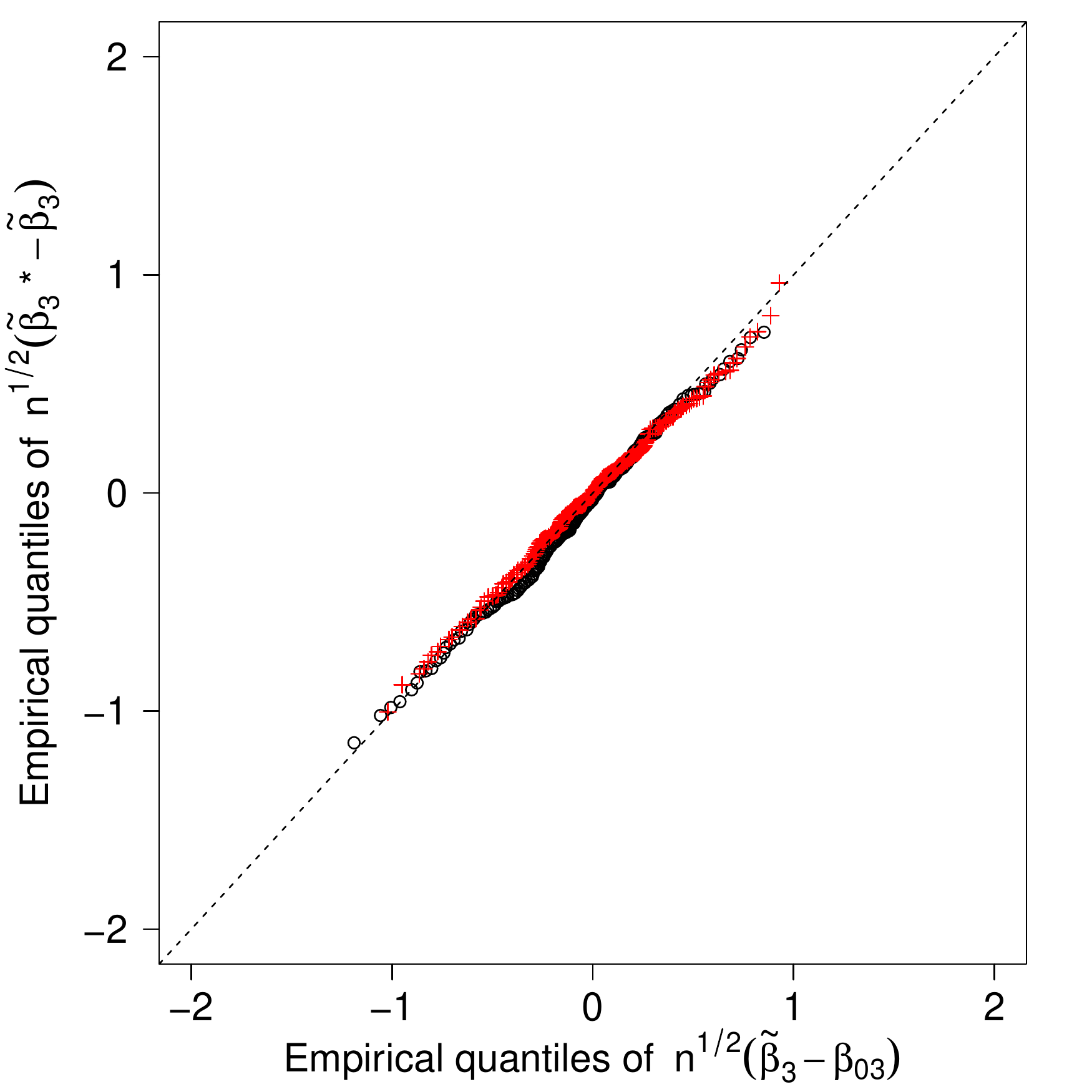} \\
~~~~~~~(c) & ~~~~~~~(d) \\[-27pt]
 \includegraphics[width=65mm]{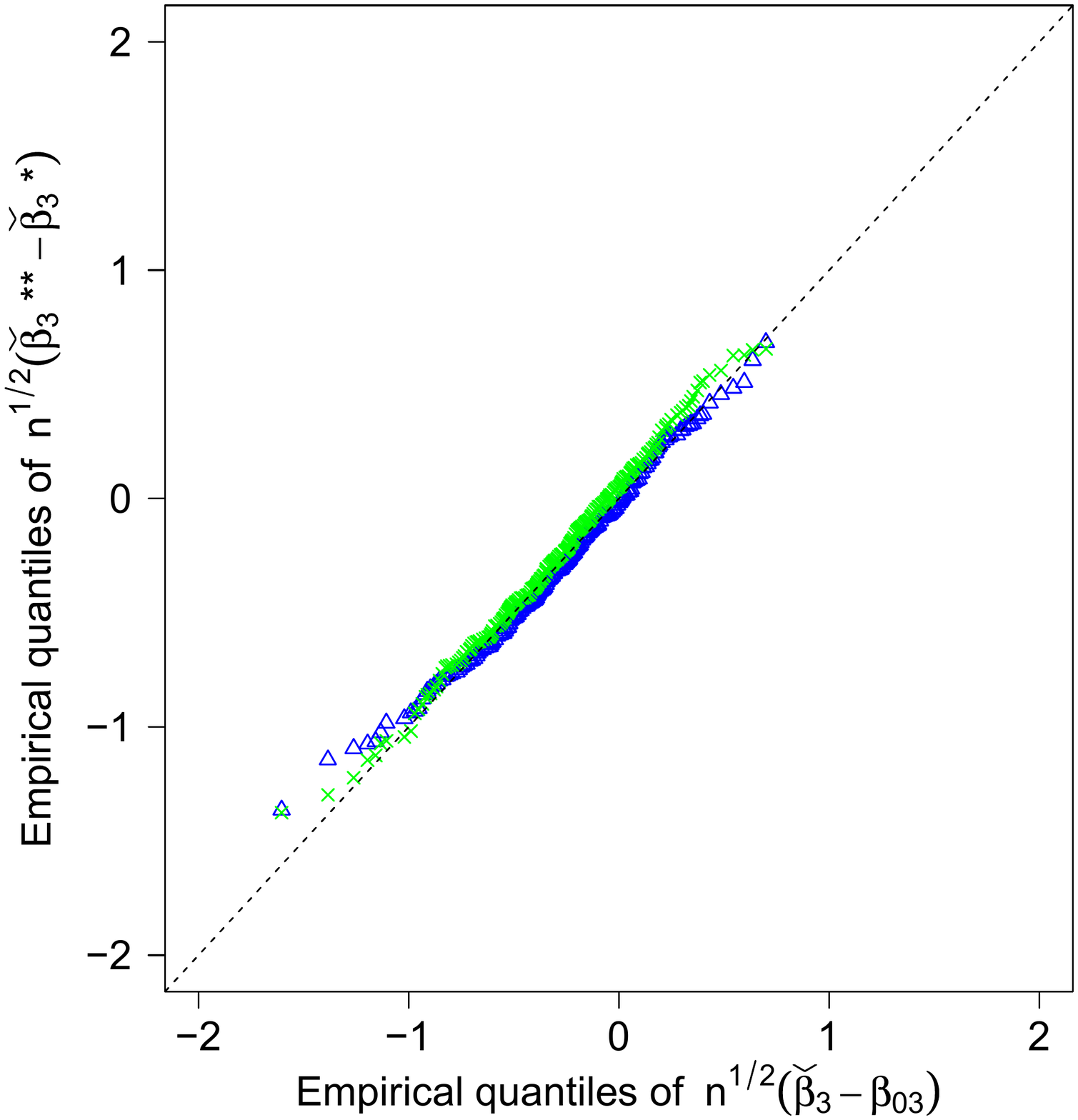} &   \includegraphics[width=65mm]{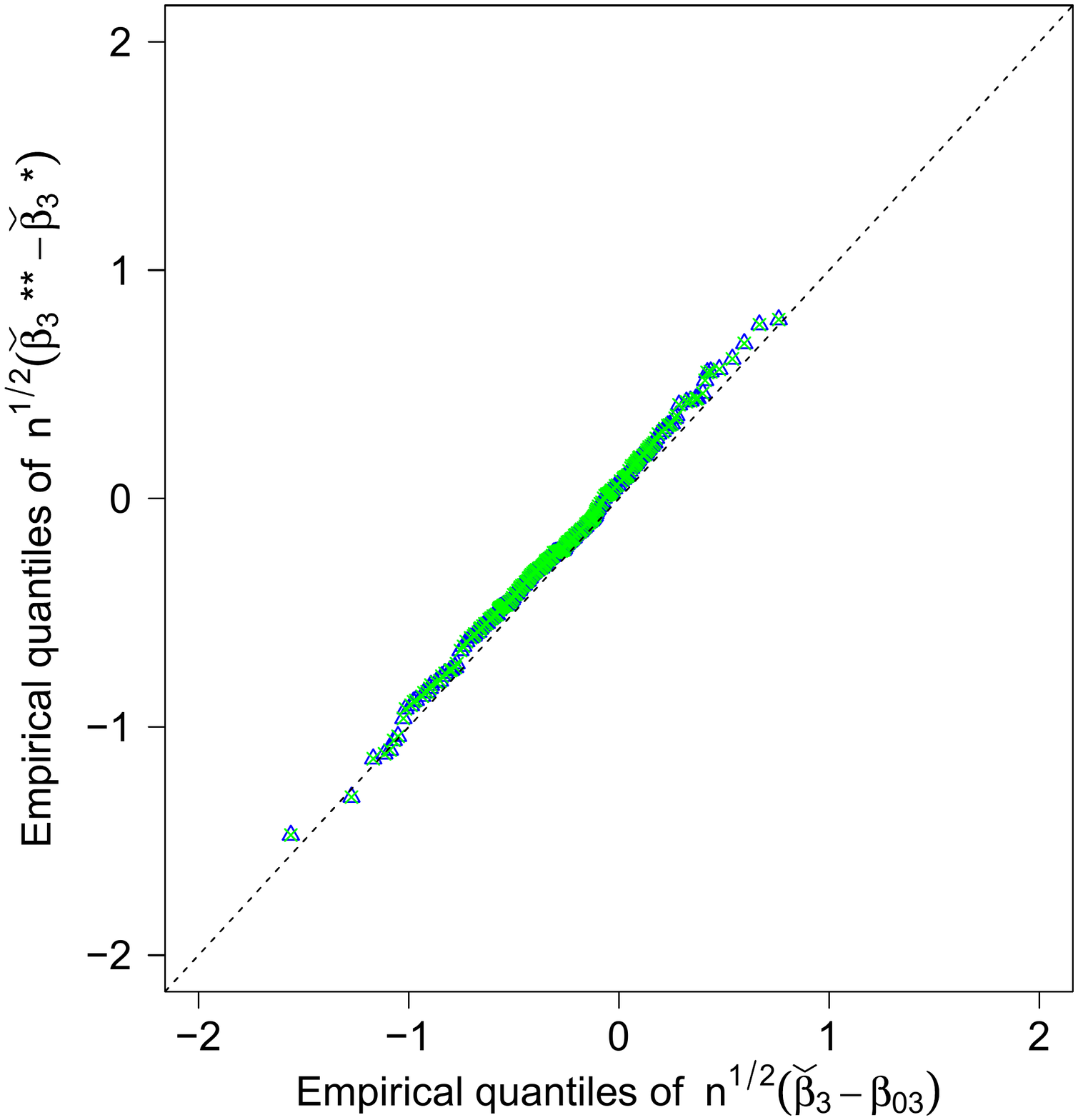} \\[-27pt]
\end{tabular}
\caption{QQ plots for the New AL1 ({\color{black} \protect\symbolALg}), New AL2 ({\color{red} \protect\symbolALgg}), New L1 ({\color{blue} \protect\symbolLa}), and New L2 ({\color{green} \protect\symbolLn}) methods for estimating 
$\beta_3=$0$\cdot$25 when $n=250$. 
(a) and (b) adaptive $L_1$ method when $\tau=$0$\cdot$5 and $\tau=$0$\cdot$7, respectively; 
(c) and (d) $L_1$ method when $\tau=$0$\cdot$5 and $\tau=$0$\cdot$7, respectively.}
\end{figure}


\end{document}